\documentclass{aa}  

\usepackage[varg]{txfonts}
\usepackage{amssymb}
\usepackage{graphicx}
\newcommand\degree{^{\circ}}
\usepackage{graphicx}	
\usepackage{mathtools}	
\usepackage{amssymb}	
\usepackage{multicol}        
\usepackage{bm}		
\usepackage{pdflscape}	
\usepackage{xcolor}
\usepackage{placeins}
\usepackage{txfonts}
\begin{document}


   \title{Supernova remnants of red supergiants: from barrels to Cygnus-loops}
   \titlerunning{From barrels to Cygnus-loops}
   \authorrunning{D. M.-A.~Meyer}

   \author{D. M.-A.~Meyer\inst{1}, 
           P.~F.~Velázquez\inst{2,6},
           M. Pohl\inst{3},
           K. Egberts\inst{1}, 
           M.~Petrov\inst{5},    
           M.~A.~Villagran\inst{6},  
           D.~F.~Torres\inst{1,7,8}, 
           R.~Batzofin\inst{1}
          \fnmsep
          }

\institute{
       $^1$ Institute of Space Sciences (ICE, CSIC), Campus UAB, Carrer de Can Magrans s/n, 08193 Barcelona, Spain  \\       
       \email{dmameyer.astro@gmail.com}\\ 
       $^{2}$ Instituto de Ciencias Nucleares, Universidad Nacional 
              Aut\'onoma de M\'exico, Ap. 70-543, CDMX, 04510, M\'exico \\       
       $^{3}$ Universit\" at Potsdam, Institut f\" ur Physik und Astronomie, 
              Karl-Liebknecht-Strasse 24/25, 14476 Potsdam, Germany \\              
       $^4$ Deutsches Elektronen-Synchrotron DESY, Platanenallee 6, 15738 Zeuthen, Germany \\
       $^{5}$ Max Planck Computing and Data Facility (MPCDF), Gießenbachstrasse 2, D-85748 Garching, Germany\\         
       $^{6}$ Instituto de Astronom\'{i}a y F\'{i}sica del Espacio (IAFE), Av. Int. G\"uiraldes 2620, 
               Pabellón IAFE, Ciudad Universitaria, 1428, Buenos Aires, Argentina \\   
       $^7$ Institut d’Estudis Espacials de Catalunya (IEEC), Gran Capità 2-4, 08034 Barcelona, Spain \\
       $^8$ Institució Catalana de Recerca i Estudis Avançats (ICREA), 08010 Barcelona, Spain \\  
}

\date{Received September 15, 1996; accepted March 16, 1997}
 
\abstract
{
Core-collapse supernova remnants are the nebular leftover of defunct massive stars 
which have died during a supernova explosion, mostly while undergoing the red 
supergiant phase of their evolution. The morphology and emission properties of 
those remnants are a function of the distribution of circumstellar material at the 
moment of the supernova, \textcolor{black}{the intrisic properties of the explosion, 
as well as those of the ambient medium.} 
By means of 2.5-dimensional numerical magneto-hydrodynamics simulations, we model 
the long-term evolution of supernova remnants generated by runaway rotating massive
stars moving into a magnetised interstellar medium. 
Radiative transfer calculations reveal that the projected non-thermal emission of the 
supernova remnants decreases with time, i.e. older remnants are fainter than younger 
ones. Older ($80\, \rm kyr$) supernova remnants whose progenitors were moving with space 
velocity corresponding to a Mach number $M=1$ ($v_\star=20\, \rm km\, \rm s^{-1}$) in the 
Galactic plane of the ISM ($n_{\rm ISM}=1\, \rm cm^{-3}$) are brighter in synchrotron 
than when moving with a Mach number $M=2$ ($v_\star=40\, \rm km\, \rm s^{-1}$). 
We show that runaway red supergiant progenitors first induce 
an asymmetric non-thermal $1.4\, \rm GHz$ barrel-like synchrotron supernova 
remnants (at the age of about $8\, \rm kyr$), before further evolving to adopt 
a Cygnus-loop-like shape (at about $80\, \rm kyr$). 
It is conjectured that a significative fraction of supernova remnants 
are currently in this bilateral-to-Cygnus-loop evolutionary sequence, 
and that this should be taken into account in the data interpretation 
of the forthcoming {\it Cherenkov Telescope Array (CTA)} observatory.
}


\maketitle
%


\section{Introduction}

Supernova remnants are chemically enriched nebulae made of gas and dust, left behind the explosive death of particular stellar objects, which do not eventually become a white dwarf or collapse as a black hole.  
In the case of high-mass ($\ge 8 \rm M_{\odot}$) progenitors, the mechanism producing the 
explosion is the so-called core-collapse process~\citep{woosley_araa_24_1986,woosley_rvmp_74_2002,smartt_araa_47_2009,langer_araa_50_2012}, releasing mass and energies~\citep{sukhbold_apj_821_2016} 
into the circumstellar nebula that was shaped by the spectacular stellar wind-interstellar medium (ISM) 
interaction at work prior to it~\citep{weaver_apj_218_1977,gull_apj_230_1979,chevalier_apj_344_1989,
wilkin_459_apj_1996,bear_mnras_500_2021}. 
Hence, the morphological and emission characteristics of supernova remnants 
are a function of the past evolution of their progenitor that strongly imprint their circumstellar 
medium as they later channel the propagation of the supernova shock wave as the remnant becomes 
larger and older, for both high-mass~\citep{kesteven_aa_183_1987,wang_nature_1992,vink_aa_307_1996,
uchida_pasj_61_2009} and low-mass~\citep{vink_aa_328_1997,vink_aarv_20_2012,williams_apj_770_2013,
broersen_mnras_441_2014}. 
To understand their observed non-thermal emission and to constrain the feedback of 
supernova remnants from massive stars into the ISM of the Milky Way, it is necessary to first 
understand in detail their shape. This medium can only be probed by numerical simulations.

\begin{table*}
	\centering
	\caption{
	List of numerical models. All simulations assume a rotating massive progenitor 
    of zero-age main-sequence mass $M_{\star}$ (in $\rm M_{\odot}$) at solar 
    metallicity and moving with velocity $v_{\star}$ through the warm phase 
    of the Galactic plane. 
    The table indicates the stellar evolution history in each model from the MS (main-sequence) phase to the final SN (supernova) explosion and ultimate PWN (pulsar wind) through the RSG (red supergiant) and WR (Wolf-Rayet) stages. 
	}
	\begin{tabular}{lccr}
	\hline
	${\rm {Model}}$       &    $M_{\star}$\, ($\rm M_{\odot}$)   
        &   $v_{\star}$ ($\rm\, km\, \rm s^{-1}$)        
	    &    Evolution history     \\ 
	\hline   
Run-20-MHD-20-SNR            &  20     &  20   &   
       $ \rm MS \rightarrow \rm RSG \rightarrow \rm SN$    \\
Run-20-MHD-40-SNR            &  20     &  40   &   
       $\rm MS \rightarrow \rm RSG \rightarrow \rm SN $     \\
Run-35-MHD-20-SNR$^{\rm a}$   &  35     &  20   &   
       $ \rm MS \rightarrow \rm RSG \rightarrow \rm WR \rightarrow \rm SN$ \\
Run-35-MHD-40-SNR$^{\rm a} $  &  35     &  40   &   
        $\rm MS \rightarrow \rm RSG \rightarrow \rm WR \rightarrow \rm SN$ \\
	\hline    
    \footnotesize 
    (a)~\citet{meyer_mnras_521_2023}
	\end{tabular}
\label{tab:table1}
\end{table*}

The realistic modelling of global supernova remnants through (magneto-)hydrodynamical 
simulations is, therefore a two-step procedure: first, the wind-ISM interaction must be 
calculated, and, secondly, the supernova explosion must be realised in it and their 
interaction subsequently simulated. 
Examples of solutions for supernova remnants of low-mass progenitors can be found in 
the studies of~\citet{orlando_aa_470_2007,petruk_393_mnras_2009,vigh_apj_727_2011,chiotellis_aa_537_2012,
chiotellis_mnras_435_2013,chiotellis_galax_8_2020,chiotellis_mnras_502_2021}. 
Simulations of remnants of high-mass progenitors are presented in the studies of~\citet{orlando_aa_470_2007,vanmarle_mnras_407_2010,orlando_apj_749_2012,
orlando_aa_622_2019,orlando_aa_636_2020,orlando_aa_645_2021}. 
Among the many degrees of freedom of the parameter space that is controlling the morphologies of core-collapse supernova remnants, the supersonic motion that is affecting a significant fraction of massive stars, i.e., of core-collapse progenitors, is a preponderant element to account for in the understanding of asymmetric supernova remnants~\citep{bear_mnras_468_2017,meyer_mnras_493_2020,meyer_mnras_521_2023} and its influence continues up to the pulsar wind nebulae developing in plerion, the remnants of high-mass progenitors in which a neutron star forms~\citep{meyer_mnras_521_2023}. 
Interestingly, the peculiar radio emission of barrel-like and horseshoe supernova remnants 
can be explained by invoking the coupling of the stellar wind history with the bulk motion of 
a progenitor undergoing a Wolf-Rayet phase before exploding~\citep{meyer_mnras_502_2021}.

Core-collapse supernova remnants have been observed by means of both their 
thermal and non-thermal emission and the mass of their progenitor constrained 
in~\citet{katsuda_apj_863_2018}, although their classification remains 
open~\citep{soker_apj_907_2021,soker_raa_2023,soker_raa_2023_23l1001S,
shishkin_mnras_522_2023}. 
This study shows that an exploding red supergiant star might generate most of such remnants. These stars shape the stellar surroundings at the pre-supernova time in a particular manner, which is a key question to understand \citep{soker_apj_906_2021}.  
A noticeable example is the Cygnus-Loop nebulae, whose peculiar morphology may arise from the interaction of a supernova shock wave with the walls of a 
cavity carved by the stellar wind of a defunct runaway red supergiant star~\citep{fang_mnras_464_2017}. 
A deeper understanding of the internal functioning of such remnants requires an additional numerical effort, bringing together higher spatial resolution, stellar rotation, and magnetisation of the cold pre-supernova wind, as well as a better post-processing of the 
results to extract synchrotron emission maps to be directly compared with real observations. 
This is of great interest, especially for the 
{\it Cherenkov Telescope Array}, 
see~\citet{2023MNRAS.523.5353A,2023APh...15002850A,2023arXiv230903712C}.

In this paper, we continue our numerical exploration of the morphologies and 
non-thermal synchrotron emission properties of core-collapse supernova remnants 
of runaway massive stars, that has been initiated in~\citet{meyer_mnras_521_2023}
by means of 2.5-dimensional magneto-hydrodynamical simulations. 
This paper focuses on the comparison between supernova remnants  of zero-age 
main-sequence $20\, \rm M_\odot$ exploding while undergoing a red supergiant, and 
compared to supernova remnants  of zero-age main-sequence $35\, \rm M_\odot$ finishing 
their life as a Wolf-Rayet star. 
We generate $1.4\, \rm GHz$ synchrotron emission maps using the method developed 
in~\citet{2023arXiv230916410V}, which allows recovery of the physical units of 
the projected emissions that are often presented in a normalised 
fashion. It is, therefore, possible to compare the brightnesses of the modelled 
supernova remnants as a function of time and of the initial conditions of the simulations. 
The synthetic emission maps are compared and discussed with real observations of 
supernova remnants.

The paper is organised as follows. In Section~\ref{method}, we introduce the reader to 
the numerical methods used to generate the results detailed in Section~\ref{results}. 
We further discuss the results in this study in Section~\ref{discussion} and conclude 
in Section~\ref{conclusion}.


\begin{figure*}
        \centering
        \includegraphics[width=0.95\textwidth]{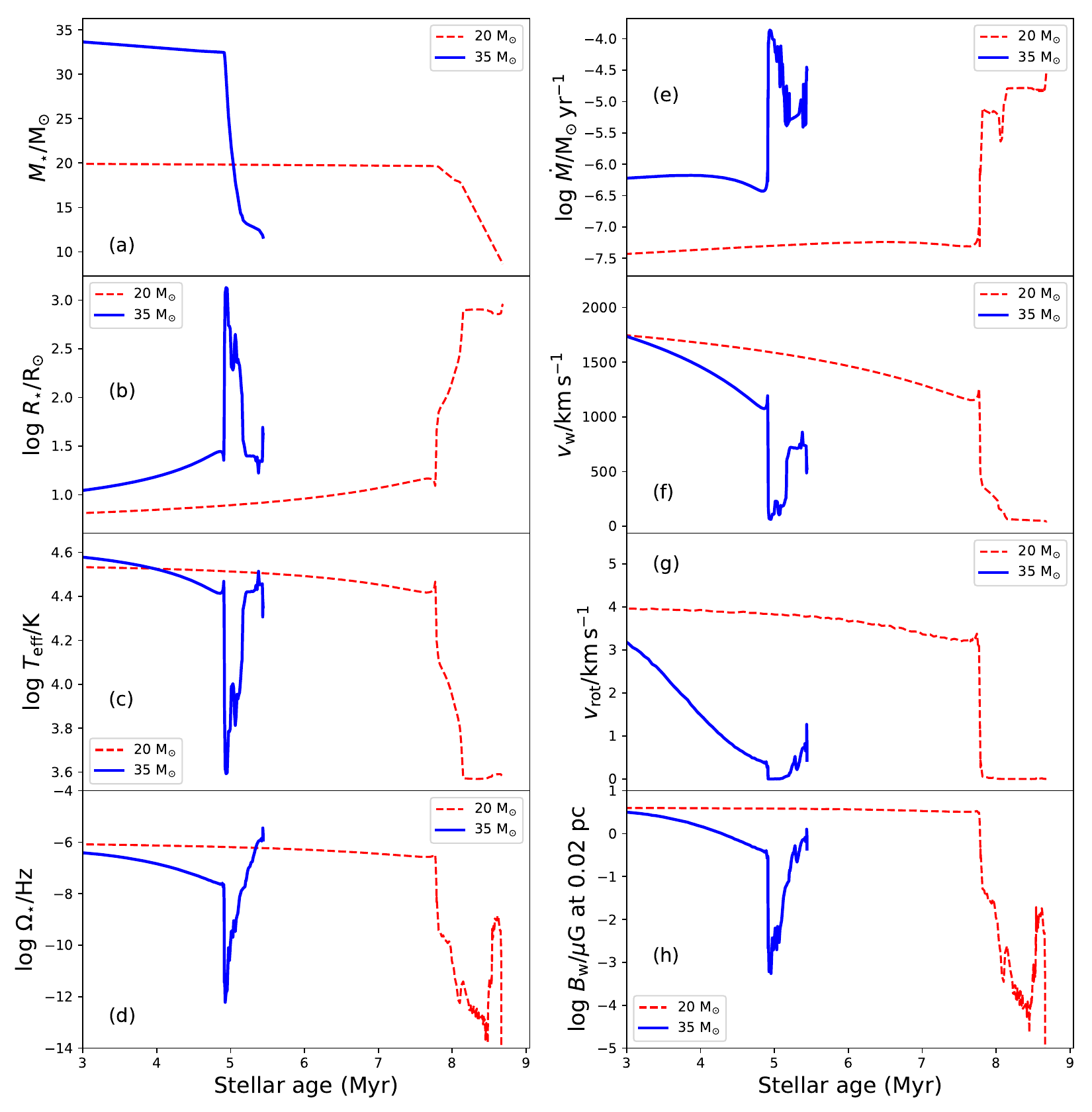}  \\
        \caption{
        Temporal evolution (in $\rm Myr$) of the supernova progenitors of 
        zero-age main-sequence $20\, \rm M_{\odot}$ (dotted red line) 
        and $35\, \rm M_{\odot}$ (solid blue line) considered in our study. 
        The panels display the stellar mass $M_{\star}$ (panel a, in $\rm M_{\odot}$), 
        radius $R_{\star}$ (panel b, in $\rm R_{\odot}$), 
        effective temperature $T_{\rm eff}$ (panel c, in $\rm K$), 
        angular frequency at the surface $\Omega_{\star}$ (panel d, in $\rm Hz$),  
        mass-loss rate $\dot{M}$ (panel e, in $\rm M_{\odot}\, \rm yr^{-1}$), 
        wind velocity $v_{\rm w}$ (panel f, in $\rm km\ \rm s^{-1}$), 
        equatorial rotation velocity $v_{\rm rot}$ (panel g, in $\rm km\ \rm s^{-1}$),  
        the magnetic field in the wind at $0.02\, \rm pc$ $B_{\star}$ (panel h, in $\mu \rm G$). 
        }
        \label{fig:stellar_properties}  
\end{figure*}

\section{Method}
\label{method}

In this section, we present the boundary and initial conditions of the magneto-hydrodynamical 
simulations performed in this study. It details the stellar evolution models utilised, the 
used calculation strategy, numerical methods and the assumptions adopted for the 
radiative transfer calculations in order to compare our results with real observations.

\subsection{Interstellar medium}
\label{method_ism}

The 
assumed ISM 
number density is taken as 
$n_{\rm ISM}=0.79\, \rm cm^{-3}$, which corresponds to that of the warm phase of the 
Galactic plane~\citep{wolfire_apj_587_2003}. The gas is assumed to be ionized and 
has a temperature of $T_{\, \rm ISM}=8000\, \rm K$~\citep{meyer_mnras_521_2023}. 
Its magnetic field of strength $B_{\rm ISM}=7\, {\rm \mu G}$~\citep{meyer_mnras_527_2024} 
is considered as uniform and organised linearly, parallel to the direction of motion of 
the star. This configuration is imposed by the two-dimensional nature of the calculations, 
see~\citet{vanmarle_584_aa_2015,meyer_mnras_464_2017}. 
The gas is initially in equilibrium 
between the cooling of the material at high temperature and the heating provided by the 
starlight, which ionises the circumstellar gas.

\textcolor{black}{
The heating rate $\Gamma$ stands for the recombining hydrogenoic ions that are 
ionized by photospheric photons. The liberated electrons get the quantum of energy 
transported by the ionizing photon. 
This process happens at a rate that is taken from the recombination coefficient  $\alpha_{\mathrm{rr}}^{\mathrm{B}}$, interpolated from the table 
4.4 of~\citet{osterbrock_1989}. 
}
The cooling term $\Lambda$ includes contributions from H, He and metals at solar 
helium abundance $Z$~\citep{wiersma_mnras_393_2009,asplund_araa_47_2009}, modified 
to include the H recombination line cooling by use of the energy loss coefficient 
case B of~\citet{hummer_mnras_268_1994} and the O and C forbidden lines by collisionally 
excited emission presented in~\citet{henney_mnras_398_2009}. 
Our study uses species 
abundance at $\rm O/\rm H=4.89\times 10^{-4}$ in number density~\citep{asplund_araa_47_2009}.

\subsection{Magnetized, rotating stellar wind}
\label{method_sw}

The stellar wind terminal radial velocity is calculated as, 
\begin{equation}    
     v_{\rm w}(t) = \sqrt{ \beta(T)  \frac{  2 G M_{\star}(t) }{ R_{\star}(t)} },
\end{equation}
where $G$ is the gravitational constant, $M_{\star}$ the stellar mass and $R_{\star}$ 
the stellar radius, respectively. The factor $\beta(T)$ is a piece-wise function of 
the temperature that is taken from the study of~\citet{eldridge_mnras_367_2006}.  
\textcolor{black}{
In this study, as in the former work of this series~\citep{meyer_mnras_521_2023}, 
the escape wind velocity is calculated using the photospheric luminosity and the effective temperature to compute the stellar radius and, therefore, wind terminal speed. 
This reduces values for the Wolf-Rayet pĥase (at $5.33\, \rm Myr$ after the zero-age main sequence). The physics for the terminal wind velocity of WR stars is not well understood, and our values are still largely in accordance with observations 
of weak-winded ($<1000\, \rm km\, \rm s^{-1}$) Galactic Wolf-Rayet stars e.g. 
WR16, WR40, WR105~\citep{hamman_aa_625_2019}. 
}
The radial-dependence of the stellar wind density reads,  
\begin{equation}
	\rho_{w}(r,t) = \frac{ \dot{M}(t) }{ 4\pi r^{2} v_{\rm w}(t) }, 
    \label{eq:wind}
\end{equation}
with $\dot{M}$ the mass-loss rate of the massive star at a given time $t$.

We consider massive stars, which, at the onset of their main sequence, initially 
rotate such that, 
\begin{equation}    
    \frac{  \Omega_{\star}(t=0) }{   \Omega_{\rm K}    } = 0.1, 
    \label{XXXXXXXXX}
\end{equation}
and with,
\begin{equation}    
     \Omega_{\star}(t) = \frac{ v_{\rm rot}(t) }{ R_{\star}(t) }, 
\end{equation}
where $ \Omega_{\star}$ and $\Omega_{\rm K}$ are the equatorial and Keplerian 
rotational velocities, respectively, and $v_{\rm rot}$ the toroidal 
velocity. The latitude-dependent toroidal component of the rotation surface 
is, therefore, 
\begin{equation}
	v_{\phi}(\theta,t) = v_{\rm rot}(t) \sin( \theta ),
\label{eq:Vphi}
\end{equation}
while its polar component is set to $v_{\theta}=0$, see also the 2.5-dimensional 
approach developed for the modelled magnetised wind of rotating stars in 
~\citet{parker_paj_128_1958,weber_apj_148_1967,pogolerov_aa_321_1997,
pogolerov_aa_354_2000,chevalier_apj_421_1994,rozyczka_apj_469_1996}.

\begin{figure}
        \centering
        \begin{minipage}[b]{ 0.475\textwidth} 
                \includegraphics[width=1.0\textwidth]{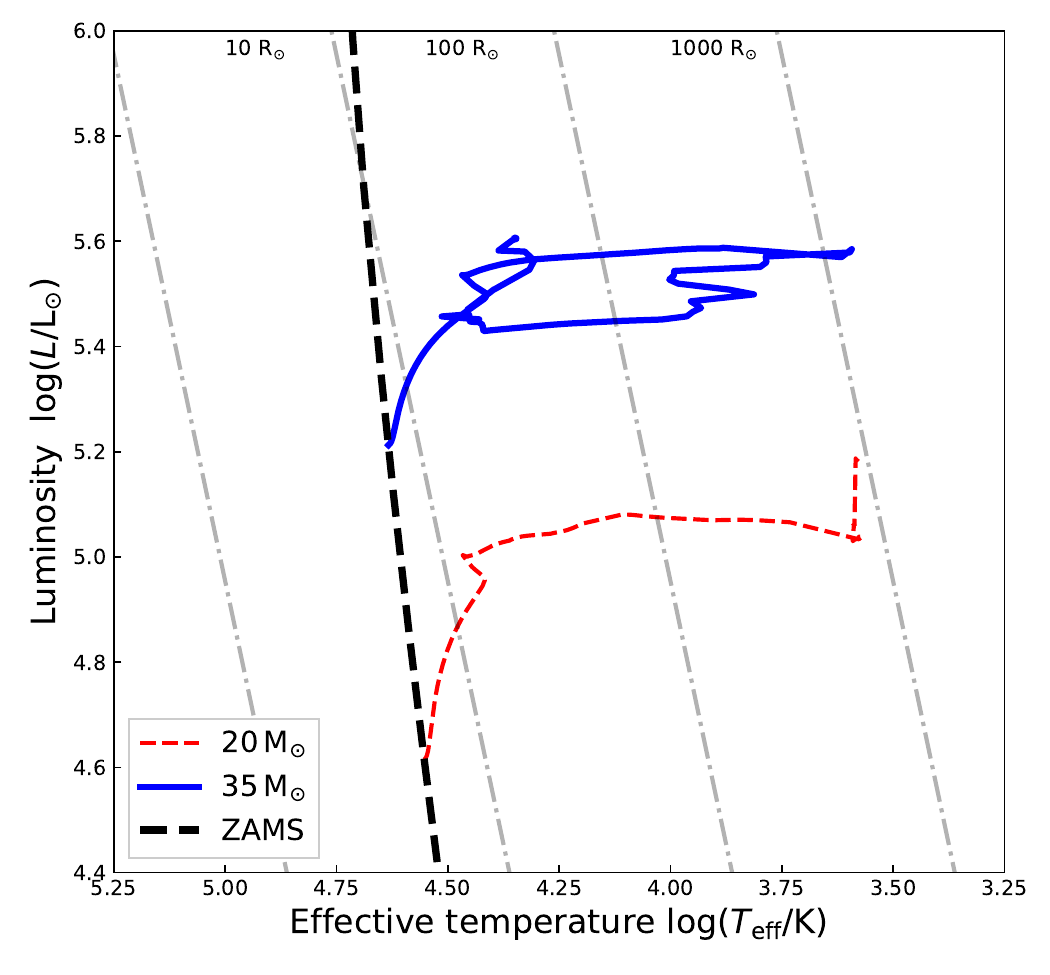}
        \end{minipage}     
        \caption{ 
        \textcolor{black}{
                Evolutionary tracks in the Hertzsprung-Russell-diagram of the 
                two progenitor stars of initial masses $20 \, \rm M_{\odot}$ 
                (dotted red) and $35 \,  \rm M_{\odot}$ (solid blue) considered 
                in this study~\citep{ekstroem_aa_537_2012}.
                }
                 }      
        \label{Fig_HRD}  
\end{figure}

The other characteristic stellar parameters are interpolated from the tabulated evolution 
models from the {\sc geneva} 
library\footnote{https://www.unige.ch/sciences/astro/evolution/en/database/syclist/}
~\citep{2008Ap&SS.316...43E,ekstroem_aa_537_2012}, which provides the evolutionary
structure and effective surface properties of high-mass stars from their zero-age 
main-sequence mass to their pre-supernova time when the Si burning phases happen. 
\textcolor{black}{
Our model, therefore, assumes that the mass-loss rate and the radial component of the wind velocity of the progenitor star are spherically symmetric throughout its entire pre-supernova evolution. Such an approach is typical and finds its origins in the early developments of stellar evolution codes~\citep{heger_apj_apj_2000,2008Ap&SS.316...43E,brott_aa_530_2011a,
brott_aa_530_2011b,ekstroem_aa_537_2012}, treating the hydrodynamic 
stellar structure equations for massive stars in a 1D-dimensional fashion as 
long they do not rotate close to their critical angular 
velocity~\citep{Georgy_a__527_2011}, 
and it is supported by observations of massive star's stellar wind close to 
the termination shock of their pc-scale nebulae~\citep{weaver_apj_218_1977,
buren_apj_329_1988,wilkin_459_apj_1996,peri_aa_538_2012,peri_aa_578_2015}, also 
leading to a spherically symmetric theory for hot-star wind acceleration 
driven by line-emission and radiation 
pressure~\citep{1980ApJ...242.1183A,1982ApJ...259..282A,1986ApJ...311..701F}. 
However, many factors, such as binarity, affecting up to $70\%$ of massive stars, impose azimuthal deviations to these winds, especially in the region in the vicinity 
of the orbital plane of massive multiple system~\citep{sana_sci_337_2012}. 
Additionally, the alternation of high and slow winds at the phase 
transitions like the onset of the red supergiant, 
blue supergiant or Wolf-Rayet evolutionary sequences are also a source of 
asymmetry~\citep{parkin_apj_726_2011,Madura_mnras_436_2013,Gvaramadze_mnras_454_2015,martayan_aa_587_2016,elmellah_aa_637_2020}. 
Since we concentrate on the large-scale surroundings of single, slowly-rotating 
objects, the spherically symmetric assumption for stellar winds is acceptable. 
For completeness, the Hertzsprung-Russel diagram of the two progenitor stars we consider in this study is plotted in Fig.~\ref{Fig_HRD}. 
}

The stellar magnetic field structure is assumed to be a Parker spiral, 
\begin{equation}
	B_{\rm r}(r,t) = B_{\star}(t) \Big( \frac{R_{\star}(t)}{r} \Big)^{2},
    \label{eq:Br}
\end{equation}
\begin{equation}
	B_{\phi}(r,t) = B_{\rm r}(r,t) 
	\left( \frac{ v_{\phi}(\theta,t) }{ v_{\rm w}(t) } \right) 
	\left( \frac{ r }{ R_{\star}(t) }-1 \right),
    \label{eq:Bphi}
\end{equation}
and
\begin{equation}
    B_{\theta}(r,t) =0
    \label{eq:Btheta}
\end{equation}
respectively. 
The total stellar wind magnetic field therefore reads, 
\begin{equation}
	B_{\rm w}(r,t) = \sqrt{ B_{\rm r}(r,t)^2 + B_{\phi}(r,t)^2 + B_{\theta}(r,t)^2 }.
    \label{eq:Bphi}
\end{equation}
The magnetic field strength at the surface of the star is scaled 
to that of the Sun, as described in~\citet{scherer_mnras_493_2020,herbst_apj_897_2020,
baalmann_aa_634_2020,baalmann_aa_650_2021,meyer_mnras_506_2021}. 
We adopt the time-dependent evolution of the surface magnetic field $B_{\star}$ of 
the supernova progenitor as derived in~\citet{meyer_mnras_521_2023,meyer_mnras_527_2024}.  
The main-sequence surface magnetic field strength is taken to 
$B_{\star}=500\, \rm G$~\citep{fossati_aa_574_2015,castro_aa_581_2015, przybilla_aa_587_2016,castro_aa_597_2017}, 
the red supergiant phase one to that of 
Betelgeuse, i.e. $B_{\star}=0.2\, \rm G$~\citep{vlemmings_aa_394_2002,vlemmings_aa_434_2005, 
kervella_aa_609_2018}, while for the Wolf-Rayet phase we assume 
$B_{\star}=100\, \rm G$~\citep{hubrig_mnras_458_2016,chevrotiere_apj_781_2014,
meyer_mnras_507_2021}.

\subsection{Supernova ejecta}
\label{method_snr}

The supernova blastwave, whose properties are determined by 
the mass of the ejecta $M_{\rm ej}$ and energy of the 
explosion $E_{\rm ej}$~\citep{shishkin_mnras_522_2023}, is modelled using 
the method of~\citet{truelove_apjs_120_1999}. The blastwave is injected 
at the time of the explosion considering the following radial density profile, 
\begin{equation}
\rho(r) = \begin{cases}
        \rho_{\rm core}(r) & \text{if $r \le r_{\rm core}$},               \\
        \rho_{\rm max}(r)  & \text{if $r_{\rm core} < r < r_{\rm max}$},    \\
        \end{cases}
	\label{cases}
\end{equation}
with an inner constant region, the so-called core, characterised by a density,  
\begin{equation}
   \rho_{\rm core}(r) =  \frac{1}{ 4 \pi n } \frac{ (10 E_{\rm ej}^{n-5})^{-3/2}
 }{  (3 M_{\rm ej}^{n-3})^{-5/2}  } \frac{ 1}{t_{\rm max}^{3} },
   \label{sn:density_1}
\end{equation}
and an outer steeply decreasing region of density
\begin{equation}
   \rho_{\rm max}(r) =  \frac{1}{ 4 \pi n } \frac{ \left(10 E_{\rm
ej}^{n-5}\right)^{(n-3)/2}  }{  \left(3 M_{\rm ej}^{n-3}\right)^{(n-5)/2}  } 
\frac{ 1}{t_{\rm max}^{3} } 
\bigg(\frac{r}{t_{\rm max}}\bigg)^{-n},
   \label{sn:density_2}
\end{equation}
where the exponent $n=11$ is typical for massive progenitor 
stars~\citep{chevalier_apj_344_1989} and where $r_{\rm max}$ is the outer 
limit of the blastwave at that time. The age of the blastwave is determined 
using the iterative procedure of~\citet{whalen_apj_682_2008}, 
such that, 
\begin{equation}
    t_{\rm max} = \frac{r_{\rm max}}{v_{\rm max}},
\end{equation}
with $v_{\rm max}=3 \times 10^{4}\, \rm km\, \rm s^{-1}$. 

The velocity profile of the blastwave is taken as, 
\begin{equation}
    v(r) = \frac{ r }{ t },
\end{equation}
which ensures that the early blastwave propagates homologously in the pre-supernova 
stellar wind of the progenitor. The ejecta velocity at a distance $r_{\rm core}$ from 
the center of the explosion reads, 
\begin{equation}
   v_{\rm core} = \bigg(  \frac{ 10(n-5)E_{\rm ej} }{ 3(n-3)M_{\rm ej} } \bigg)^{1/2}, 
   \label{sn:vcore}
\end{equation}
see ~\citet{truelove_apjs_120_1999,whalen_apj_682_2008,vanveelen_aa_50_2009,vanmarle_mnras_407_2010}. 
The mass of defunct stellar material in the supernova ejecta is determined by 
subtracting the mass of a neutron star $M_{\mathrm{NS}}=1.4\, \rm M_{\odot}$ 
to that of the progenitor at the moment of the explosion, 
\begin{equation}
   M_{\rm ej} =  M_{\star} - \int_{t_\mathrm{ZAMS}}^{t_\mathrm{SN}} \dot{M}(t)~ dt - M_{\mathrm{NS}},
   \label{eq:co}
\end{equation}
which is $7.28\, \rm M_{\odot}$ and $10.12\, \rm M_{\odot}$ for the $20\, \rm M_{\odot}$ 
and $35\, \rm M_{\odot}$ progenitors, respectively.

\textcolor{black}{
Our setting of the supernova blastwave assumes spherical symmetry of the radial component of the supernova ejecta. This is a classical approach to numerical simulations like ours~\citep{vanveelen_aa_50_2009,vanmarle_mnras_407_2010}, consisting of considering that the homologous expansion of the supernova material through the last stellar wind is not affected by any latitude- or azimuthal-dependent hydrodynamical flow. One should mention that the physics of 
core-collapse supernova explosion is complex and far beyond the simple prescription 
used in this study. Particularly the anisotropic in the emission of a released neutrino 
can greatly change the explosion energy and channel it in the direction of the neutrino 
anisotropies, producing anisotropic blastwave~\citep{shimizu_apj_552_2001,
2012A&A...537A..63M,gabler_mnras_502_2021}. 
Furthermore, the presence of a binary companion leads to additional blastwave-circumstellar 
interactions potentially producing so-called common envelope jets supernova in which a first 
component explodes, giving a compact object (black hole or neutron star) that accretes 
material from its post-main-sequence companion, such as a supergiant 
star~\citep{2011MNRAS.416.1697P,2014MNRAS.438.1027P,2016ApJ...826..178G,
2018MNRAS.478..682B,2020MNRAS.492.3013K,2022RAA....22i5007S,2022MNRAS.516.4942S,
2022RAA....22l2003S,soker_raa_2023}. 
Since our work considers single stars explosion of the canonical energy of 
$10^{51}\, \rm erg$, we neglect any anisotropy in the supernova blastwave and/or 
jet pushing out the gas of a former common envelope.  
}

\subsection{Governing equations}
\label{method_eq}

The equations solved on the grid meshing the computational domain are those of 
the non-ideal magneto-hydrodynamics. They read as, 
\begin{equation}
	   \frac{\partial \rho}{\partial t}  + 
	   \vec{ \nabla } \cdot \big(\rho\vec{v}) =   0,
\label{eq:mhdeq_1}
\end{equation}
\begin{equation}
	   \frac{\partial \vec{m} }{\partial t}  + 
           \vec{\nabla} \cdot \Big( \vec{m} \otimes \vec{v}  
           - \vec{B} \otimes \vec{B} + \vec{\hat I}p_{\rm t} \Big)  
            =   \vec{0},
\label{eq:mhdeq_2}
\end{equation}
\begin{equation}
	  \frac{\partial E }{\partial t}   + 
	  \vec{\nabla} \cdot \Big( (E+p_{\rm t})\vec{v}-\vec{B}(\vec{v}\cdot\vec{B}) \Big)  
	  = \Phi(T,\rho),
\label{eq:mhdeq_3}
\end{equation}
%
%
\begin{equation}
	  \frac{\partial \vec{B} }{\partial t}   + 
	  \vec{\nabla} \cdot \Big( \vec{v}  \textcolor{black}{\otimes} \vec{B} - \vec{B} \textcolor{black}{\otimes} \vec{v} \Big)  = \vec{0},
\label{eq:mhdeq_4}
\end{equation}
where $\rho$ is the mass density, $\vec{v}$ the vector velocity, $\vec{m}$ the 
vector momentum, and $\vec{B}$ the vector magnetic field, respectively. In the above 
relations, $\vec{\hat I}$ is the identity vector, $\vec{0}$ is the null vector, 
and $p_{\rm t}$ is the total pressure (thermal $p$ plus magnetic 
$B^{2}/8\pi$ ) of the gas. 
The total energy of the gas reads, 
\begin{equation}
	E = \frac{p}{(\gamma - 1)} + \frac{ \vec{m} \cdot \vec{m} }{2\rho} 
	    + \frac{ \vec{B} \cdot \vec{B} }{8\pi},
\label{eq:energy}
\end{equation}
and the system is closed with the relation, 
\begin{equation}
	  c_{\rm s} = \sqrt{ \frac{\gamma p}{\rho} },
\label{eq:cs}
\end{equation}
with $c_{\rm s}$ the sound speed of the gas, with $\gamma$ the adiabatic index.

The right-hand side term, 
\begin{equation}  
	  \Phi(T,\rho)  =  n_{\mathrm{H}}\Gamma(T)   
		   		 -  n^{2}_{\mathrm{H}}\Lambda(T),
\label{eq:dissipation}
\end{equation}
of Eq.\ref{eq:mhdeq_3} represents the losses and heating by optically-thin 
radiative processes. 
The H gas number density is, 
\begin{equation}    
    n_{\mathrm{H}}= \frac{ \rho }{ \mu (1+\chi_{\rm He,Z}) m_{\mathrm{H}} }, 
\end{equation}
with $\mu$ the mean molecular weight, $\chi_{\rm He,Z}$ the He and metals 
(species of atomic number larger or equal to 2) and $m_{\mathrm{H}}$ the proton 
mass, respectively, and 
\begin{equation}
	T =  \mu \frac{ m_{\mathrm{H}} }{ k_{\rm{B}} } \frac{p}{\rho},
\label{eq:temperature}
\end{equation}
is the temperature of the gas.

\subsection{Synchrotron emission calculation}
\label{method_rad}

Our study aims to investigate the radio appearance of the supernova remnants. 
To this end, let us first call the relativistic electron energy distribution, 
\begin{equation}
        N(\gamma) = K \gamma^{-p} \propto \gamma^{-p},
        \label{eq:N}  
\end{equation}
with $K$ a proportionality constant, $p$ an index function of the spectral index 
via $p=2\alpha+1$ and $\gamma$ the Lorentz factor, 
\begin{equation}
        \gamma = \frac{ E  }{  E_{0}  } =  \frac{ \gamma m_{0} c^{2}  }{  m_{0} c^{2}  },
        \label{eq:gamma}  
\end{equation}
with $m_0$ the rest mass of the electrons and $c$ the speed of light.

The synchrotron emissivity at a frequency $\nu$ is defined for a given Lorentz factor range as, 
\begin{equation}
        \epsilon_{\rm sync}(\nu,\theta) = 
        \frac{ 1  }{  4 \pi  } \int_{ \gamma_1 }^{ \gamma_2 } 
        N(\gamma) P_{\rm sync}(\nu,\gamma,\theta) d\gamma,
        \label{eq:eps_1}  
\end{equation}
with $\gamma_1 < \gamma_2$ which can be, using the Eq. 4.43 in \citet{ghiselini_book}, 
rewritten as, 
\begin{equation}
        \epsilon_{\rm sync}(\nu ) = 
        \frac{ 3 \sigma_{\rm T} c K U_{\rm B} }{  8 \pi^2 \nu_{\rm L}  } (\sin{\theta})^{(p+1)/2}
        \Big( \frac{ \nu  }{  \nu_{\rm L}  } \Big)^{  -\frac{ p-1  }{  2  }  }
        f_{\rm sync}(p), 
        \label{eq:eps_2}  
\end{equation}

\begin{figure*}
        \centering
        \includegraphics[width=0.8\textwidth]{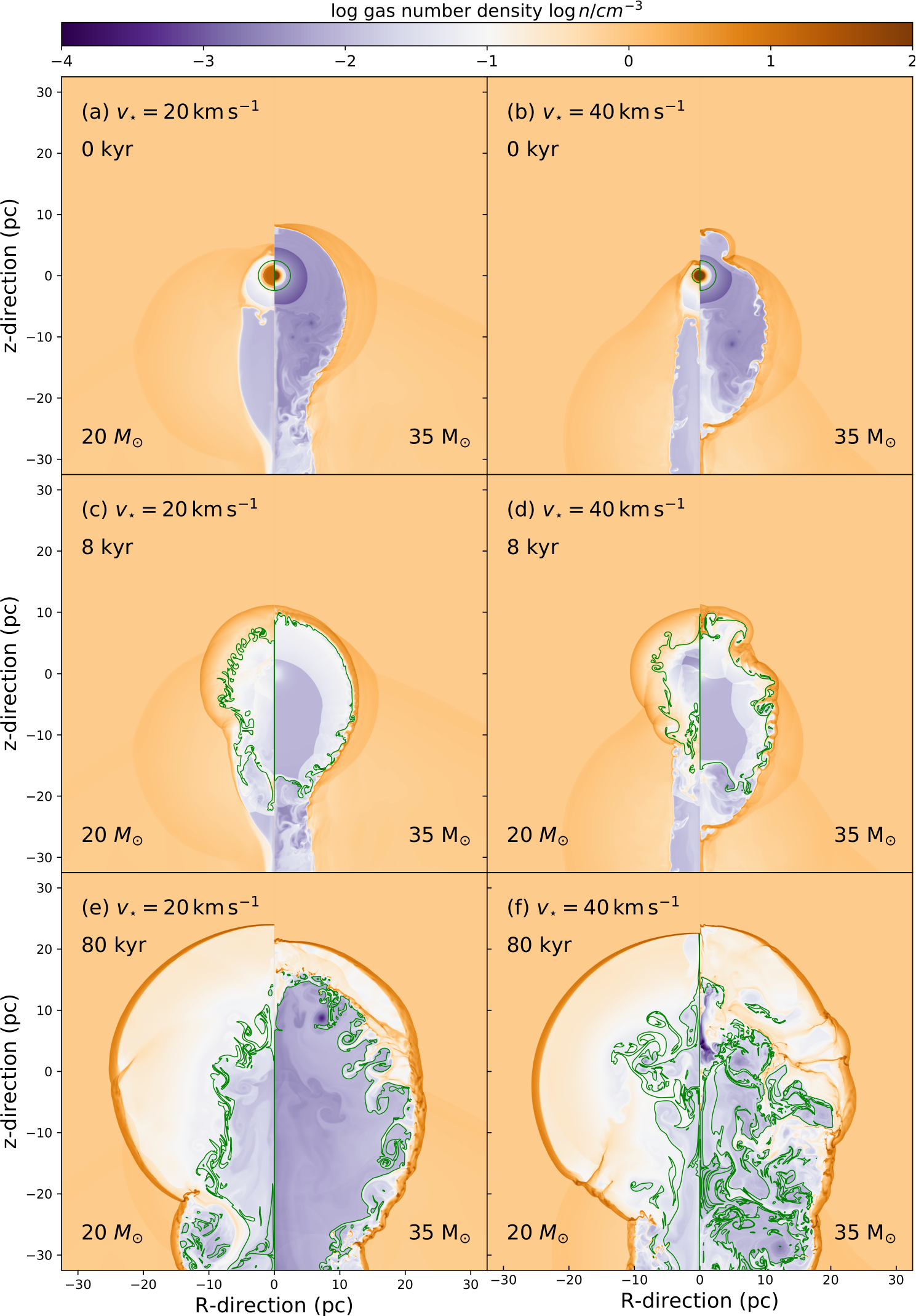}  \\
        \caption{
        Comparison between the number density fields 
        in the  supernova remnant models of a runaway 
        $20\, \rm M_{\odot}$ (left-hand part of the panels) and a 
        $35\, \rm M_{\odot}$ (right-hand part of the panels) star 
        rotating with $\Omega_{\star}/\Omega_{\rm K}=0.1$ 
        and moving with velocity 
        $v_{\star}=20\, \rm km\, \rm s^{-1}$ (left column of panels) and 
        $v_{\star}=40\, \rm km\, \rm s^{-1}$ (right column of panels). 
        The green contour mark the parts of the supernova remnants 
        with a $50\%$ contribution of ejecta. 
        }
        \label{fig:SNR_time_evolution_comparison}  
\end{figure*}

\FloatBarrier

with $\sigma_{\rm T}$ the Thomson cross section, the special function, 
\begin{equation}
        f_{\rm sync}(p) = 
        3^{ p/2 }
        \frac{ \Gamma\Big(  \frac{  3p-1 }{ 12 }   \Big)
        \Gamma\Big(    \frac{  3p+19 }{ 12 }    \Big)}
        { p + 1}  
        \label{eq:eps_3}  
\end{equation}
the magnetic density energy,  
\begin{equation}
        U_{\rm B} = \frac{  B^{ 2 } }{ 8\pi },
        \label{eq:freq_mag}  
\end{equation}
and the Larmor frequency, 
\begin{equation}
        \nu_{\rm L} = \frac{  eB }{ 2\pi m_e c },
        \label{eq:freq_larmor}  
\end{equation}
where $m_e$ is the electron mass.

The local magnetic field strength $B$ in the above relations is substituted 
by its component normal to the line-of-sight of unit vector $\vec{l}$ defining 
the viewing angle of the observer $\theta_{\rm obs} =  \angle (\vec{l},\vec{B})$. 
\textcolor{black}{
The magnetic total intensity relates to the normal component, 
\begin{equation}
        B_{\perp} =  |\vec{B} \cdot \vec{l}|
        = |\vec{B}| |\vec{l}| \sin(  \theta_{\rm obs}  )
        = |\vec{B}| \sin(  \theta_{\rm obs}  ),
        \label{eq:BBB}  
\end{equation}
which we express as,    
\begin{equation}
        B_{\perp} = |\vec{B}| 
        \sqrt{ 1 - \cos(  \theta_{\rm obs}  )^{2}  },
        \label{eq:BBB}  
\end{equation}
where, 
\begin{equation}
        |\vec{B}| = \sqrt{ B_{\rm R}^{2} + B_{\rm z}^{2} + B_{\phi}^{2} }.
        \label{eq:BB}  
\end{equation}
with $B_{\rm R}$, $B_{\rm z}$ and $B_{\phi}$ its cylindrical components. 
We need to find the value of $\cos(  \theta_{\rm obs}  )$, that is 
done via the following vectorial product, 
\begin{equation}
        |\vec{B} \times \vec{l}| 
        = |\vec{B}| |\vec{l}| \cos(  \theta_{\rm obs}  )
        = |\vec{B}| \cos(  \theta_{\rm obs}  ), 
        \label{eq:cross_product}
\end{equation}
or, 
\begin{equation}
        \cos(  \theta_{\rm obs}  )
        = \frac{ |\vec{B}\times\vec{l}| }{ |\vec{B}| }, 
        \label{eq:cos} 
\end{equation}
and, finally, combining Eqs. \ref{eq:BBB}, \ref{eq:cos} one obtains, 
\begin{equation}
B_{\perp} = |\vec{B}| \sqrt{ 1 - \Big( \frac{ |\vec{B}\times\vec{l}| }{ |\vec{B}| } \Big)^{2}  },   
\end{equation}
respectively. The modulus of the cross product (left-hand side is of 
Eq. \ref{eq:cross_product}) is calculated explicitly with the 
cylindrical components of $\vec{B}$ and $\vec{l}$ expressed in 
the cylindrical coordinate system of the numerical simulation. 
 }

Reminding that both non-thermal electron density and non-thermal electron energy 
are linked to the plasma number and energy densities with the following relations, 
\begin{equation}
        x_{\rm n} n  
        = \int_{ \gamma_{\rm min} }^{ +\infty } K \gamma ^{-p} d\gamma
         \approx \frac{  K }{ p-1 } \gamma_{\rm min}^{-p+1}, 
        \label{eq:xn}  
\end{equation}
and,  
\begin{equation}
        x_{\rm e} \epsilon_g = \int_{\gamma_{min}}^{\infty}K \gamma^{-p} (\gamma - 1) m_e c^2 d\gamma, 
        \label{eq:xe}  
\end{equation}
\textcolor{black}{with $n$ the number density and $\epsilon_g$ the energy density 
of the gas, respectively. The quantities $x_n$ and $x_e$  are fractions depending 
on $\gamma_{\rm min}$, with $x_n>0$ being related to the injection factor of the 
accelerated particles and $x_e$ a variable related to the non-thermal electrons 
cooling. 
}
Then, one can rewrite Eqs. \ref{eq:xn}-\ref{eq:xe} as, 
\begin{eqnarray}
        x_{\rm e} \epsilon_g &=&   
        \int_{ \gamma_{\rm min} }^{ +\infty } K \gamma_e^{-p} (\gamma_e-1)  m_e c^2  d\gamma = \\
        &\simeq& m_e c^2 \underbrace{\frac{K}{p-1}\gamma_\mathrm{min}^{-p+1}}_{=\chi_n n} \bigg(\underbrace{\frac{p-1}{p-2} \gamma_\mathrm{min}}_{\gg 1}-1\bigg), 
        \label{eq:xe1}  
\end{eqnarray}
with,
\begin{equation}
        \gamma_{\rm min} =    
        \frac{   x_{\rm e} \epsilon_g }{ x_{\rm n} n } \frac{ p-2 }{ p-1 } \frac{  1 }{ m_e c^2 }, 
        \label{eq:xe2}  
\end{equation}
and with the above proportionality constant of Eq. \ref{XXXXXXXXX}, 
\begin{equation}
        K  
        = (p-1) x_{\rm n} n \gamma_{\rm min}^{p-1} , 
        \label{eq:xn1}  
\end{equation}
respectively. 


The frequency-dependent synchrotron emissivity finally takes the form of, 
\begin{equation}
        \epsilon_{\rm sync}(\nu) = 
        C_{00}
        C_{01}
         \Bigg(
        \frac{ (p-1)  x_{\rm n} n  }{  ( m_e c^2)^{p-1} } 
         \Bigg) (\sin{\theta})^{(p+1)/2}
        \nu^{  -\frac{p-1}{2} }
        f_{\rm sync}(p), 
        \label{eq:eps_2}  
\end{equation}
with the coefficients, 
\begin{equation}
        C_{00} = 
        \frac{ 3  }{  8  } 
        \frac{ \sigma_{\rm T} c }{  \pi^2  }
        c
        \frac{ B^2 }{  8 \pi  }
        \Big( \frac{  eB }{ 2\pi m_e c }  \Big)^{\frac{p-3}{2}},
        \label{eq:c1}  
\end{equation}
and, 
\begin{equation}
        C_{01} = 
        \Big(  \frac{   x_{\rm e} }{ x_{\rm n} } \frac{ p-2 }{ p-2 } \frac{ \epsilon_g}{ n m_e c^2} \Big)^{p-1},
        \label{eq:c1}  
\end{equation}
respectively, \textcolor{black}{which we calculate assuming that 
$x_e$=$x_n$=0.1 and $p=2.2$ as in \citet{2023arXiv230916410V} }. 

The intensity emission map at a given frequency $\nu$ is finally generated by 
projecting the non-thermal emissivity, 
\begin{equation}
        I_{\rm sync}( \nu ) 
        =\int \epsilon_{\rm sync} ( \theta_{\rm obs } )    dl,
        \label{eq:intensity}  
\end{equation}
to obtain an emission map with an aspect angle $\theta_{\rm obs}$
between the observer's line of sight and the plane into 
which the supernova remnant lie.

\begin{figure*}
        \centering
        \includegraphics[width=0.69\textwidth]{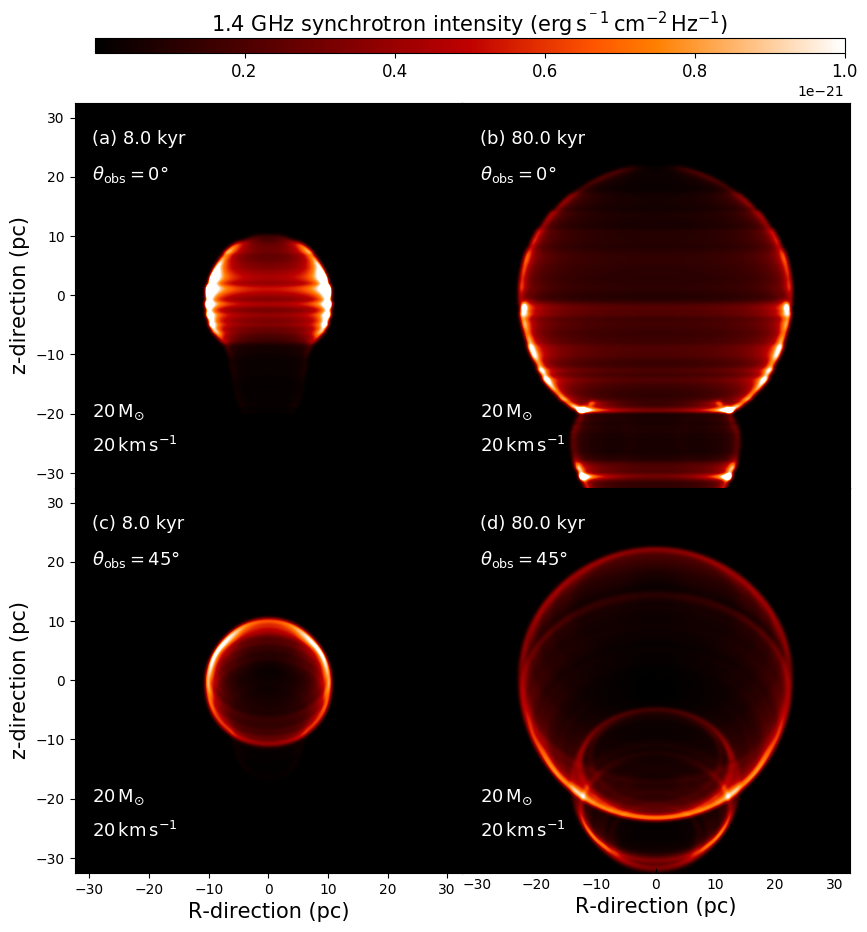}     \\         
        \caption{
        Non-thermal $1.4\, \rm GHz$ synchrotron emission maps
        (in $\rm erg\, \rm s^{-1}\, \rm cm^{-2}\, \rm sr^{-1}, \rm Hz^{-1}$) 
        for the supernova remnant generated by a $M_\star = 20\, \rm M_{\odot}$ 
        progenitor moving at $v_{\star}=20\, \rm km\, \rm s^{-1}$, respectively.  
        The remnants are shown are times $8\, \rm kyr$ 
        (\textcolor{black}{left} panels) and $80\, \rm kyr$ 
        (\textcolor{black}{right} panels) after the explosion, 
        with a viewing angle of the observer that is inclined by 
        $\theta = 0\degree$ (\textcolor{black}{top} panels) and 
        $\theta = 45\degree$ (\textcolor{black}{bottom} panels) 
        with respect to the plane of the object. 
        \textcolor{black}{
        The ring-like structures result in the mapping of 2.5-dimensional simulations 
        being mapped into a 3-dimensional shape. 
        }
        }
        \label{fig:sync_map_20_20kms}  
\end{figure*}

\begin{figure*}
        \centering    
        \includegraphics[width=0.69\textwidth]{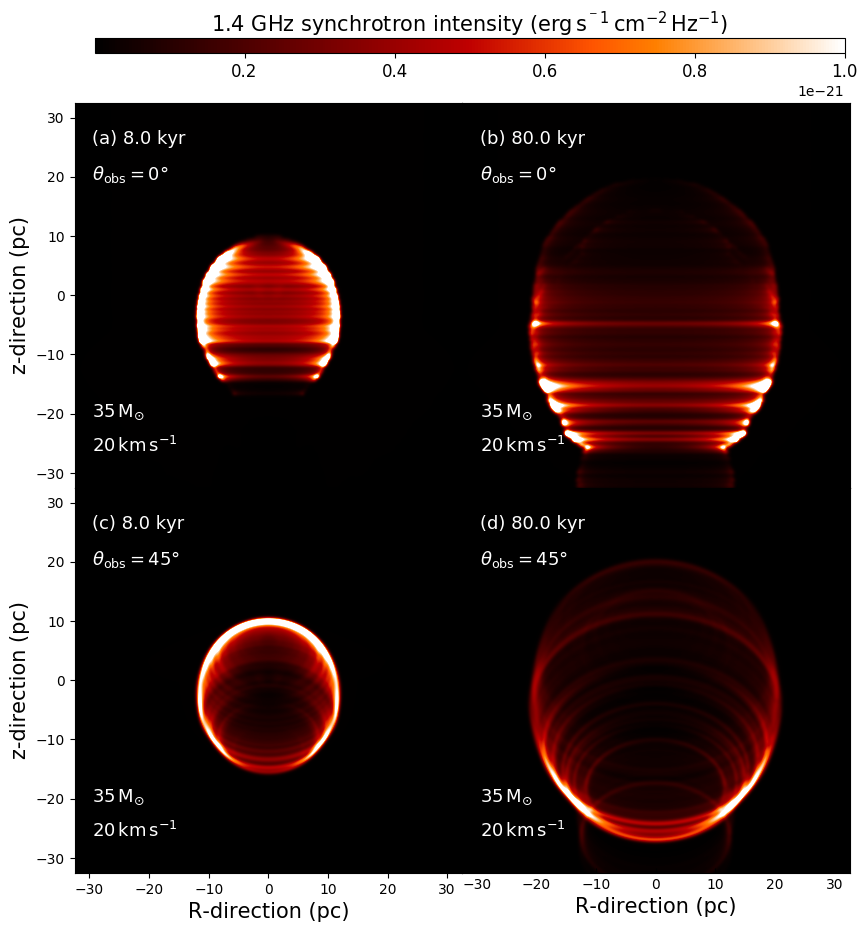}      
        \caption{
        As Fig. \ref{fig:sync_map_20_20kms} for $M_\star = 35\, \rm M_{\odot}$ progenitor 
        moving with velocity $v_{\star}=20\, \rm km\, \rm s^{-1}$.
        \textcolor{black}{
        The ring-like structures result in the mapping of 2.5-dimensional simulations 
        being mapped into a 3-dimensional shape. 
        }
        }
        \label{fig:sync_map_35_20kms}  
\end{figure*}

\begin{figure*}
        \centering
        \includegraphics[width=0.7\textwidth]{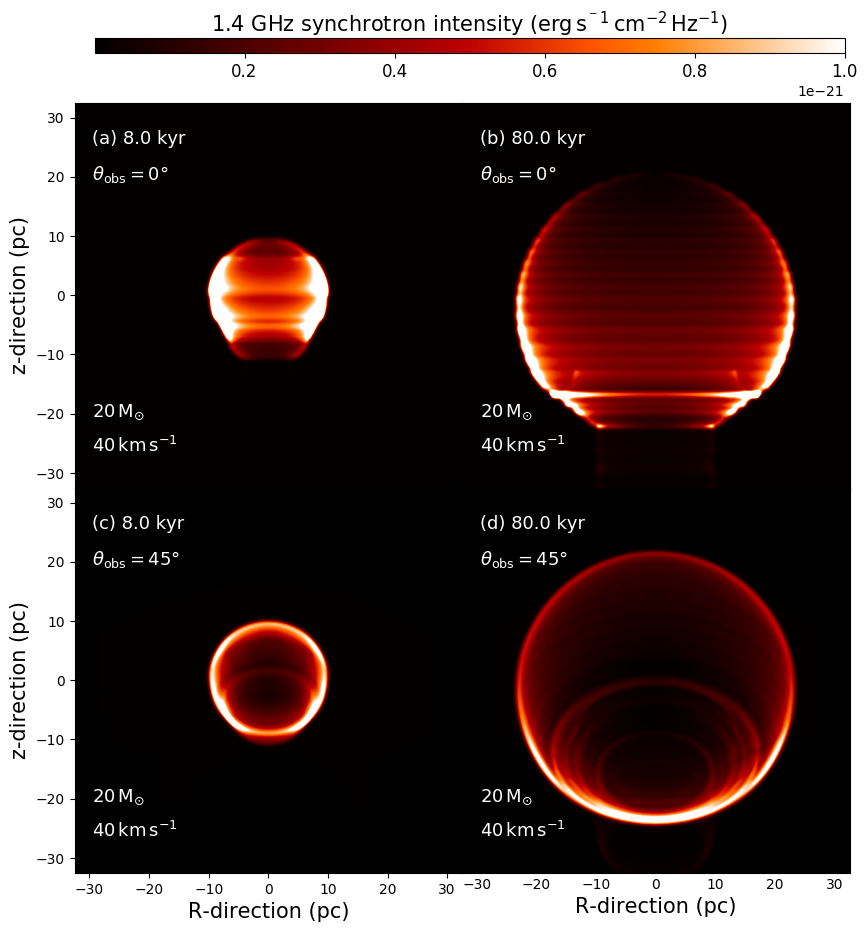}       \\        
        \caption{
        As Fig. \ref{fig:sync_map_20_20kms} for $M_\star = 20\, \rm M_{\odot}$ progenitor 
        moving with velocity $v_{\star}=40\, \rm km\, \rm s^{-1}$.
        \textcolor{black}{
        The ring-like structures result in the mapping of 2.5-dimensional simulations 
        being mapped into a 3-dimensional shape. 
        }
        }
        \label{fig:sync_map_20_40kms}  
\end{figure*}

\begin{figure*}
        \centering
        \includegraphics[width=0.7\textwidth]{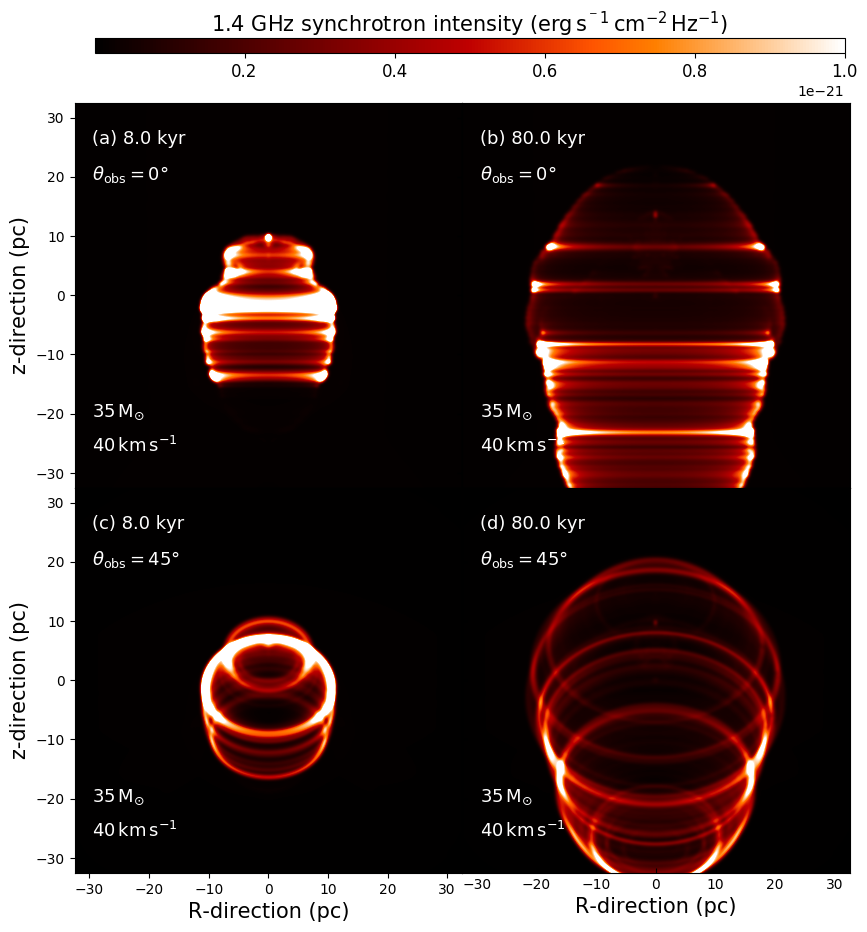}      
        \caption{
        As Fig. \ref{fig:sync_map_20_20kms} for $M_\star = 35\, \rm M_{\odot}$ progenitor 
        moving with velocity $v_{\star}=40\, \rm km\, \rm s^{-1}$.
        \textcolor{black}{
        The ring-like structures result in the mapping of 2.5-dimensional simulations 
        being mapped into a 3-dimensional shape. 
        }
        }
        \label{fig:sync_map_35_40kms}  
\end{figure*}

\subsection{Modelling strategy and numerical methods}
\label{method_strategy}

This project extends the study of~\citet{meyer_mnras_521_2023} to the regime of core-collapse 
supernova progenitors, which do not undergo a Wolf-Rayet phase but end their lives as a red supergiant star. To this end, it compares the morphologies and non-thermal appearances of their middle-aged to older ($80\ \rm kyr$) supernova remnants. The stellar wind-ISM interaction is first modelled during the entire star's life, providing the circumstellar medium's MHD structure, which is later used as initial conditions for the blastwave-surroundings interaction. 
The progenitor star wind is injected into the computational domain in a circular zone of radius $r_{\rm in}=0.01\, \rm pc$ from the origin. 
Throughout the simulations, stellar motion is taken into account by setting a velocity $v=-v_{\star}$, being $v_{\star}$ the velocity of the moving star along the direction of the 
magnetic field lines~\citet{meyer_mnras_464_2017}. A parameter space of two distinct 
velocities is explored, namely $v_{\star}=20\, \rm km\, \rm s^{-1}$ and $v_{\star}=40\, \rm km\, \rm s^{-1}$. 
We summarize the models in this work in Tab \ref{tab:table1}.

\textcolor{black}{
The ejecta-circumstellar medium is calculated in the frame of the moving star in a 2.5-dimensional cylindrical coordinate system (O;R,z) or origin $O$ 
that includes a toroidal component for each vector, rotational invariant 
with respect to the axis of symmetry $Oz$}. 
The equations are solved in a computational domain $[0;100]\times[-50;50]\, \rm pc^{2}$ mapped with an uniform mesh of $2000 \times 4000$ cells. 
The numerical simulations are performed using the {\sc pluto} 
code~\citep{mignone_apj_170_2007,migmone_apjs_198_2012,vaidya_apj_865_2018}
\footnote{http://plutocode.ph.unito.it/}, with a Godunov-type solver made of 
HLL Rieman solver~\citep{hll_ref}, a Runge-Kutta integrator and the divergence-free 
eight-wave finite-volume algorithm~\citep{Powell1997}, respectively. 
We refer the reader interested in further details on the numerical scheme and on the limitation of the method to the study of \citet{meyer_mnras_507_2021}. 
The projection of the synchrotron emissivity is perfomed using a modified 
version of the radiative transfer 
code {\sc radmc-3d}\footnote{https://www.ita.uni-heidelberg.de/$\sim$dullemond/software/radmc-3d/}, 
which performs the ray-tracing along a particular line of sight. 
The physical and characteristic quantities of the supernova progenitors are plotted in 
Fig. \ref{fig:stellar_properties}.


\section{Results}
\label{results}


\subsection{Supernova remnants morphology}
\label{resultats_morpho}

Fig. \ref{fig:SNR_time_evolution_comparison} plots the surroundings of the supernova progenitors considered in this study, at the time of onset of the blastwave propagation
and at later times, when it propagates through the wind bubble. 
In each panel, the left-hand side part refers to the model for the zero-age main sequence 
$M_\star = 20\, \rm M_{\odot}$ star, while the right-hand part concerns the 
$M_\star = 35\, \rm M_{\odot}$ progenitor star. 
The simulations are displayed for runaway stars moving with velocity 
$v_{\star}=20\, \rm km\, \rm s^{-1}$ (left column of panels) and 
$v_{\star}=40\, \rm km\, \rm s^{-1}$ (right column panels). 
They represent the number density field (in $\rm cm^{-3}$) in the simulations, 
onto which a green isocontour marks the regions of the supernova remnants that are made of equal quantities of ejecta and stellar wind material.

In both models, the circumstellar medium has the typical morphology produced around an evolving stellar wind that runs through the ISM, with a bow shock ahead of the direction of stellar motion and a cavity behind it, see Fig. 
\ref{fig:SNR_time_evolution_comparison}a,b). 
In the early phase of their evolution, the supernova remnants conserve the global appearance of their progenitor's circumstellar medium since the blastwave has not yet interacted with the stellar surroundings. They are constituted of a large, complex stellar wind bow shock, the result of the interaction of the wind blown by the star throughout the various evolutionary phases of its life~\citep{meyer_mnras_450_2015}. 
The power of stellar winds is stronger for Wolf-Rayet stars than for red supergiant progenitors. Their associated final bow shocks are wider and present large eddies due 
to Rayleigh-Tayler instabilities in the case of the $M_\star = 35\, \rm M_{\odot}$ model~\citep{brighenti_mnras_273_1995,brighenti_mnras_277_1995}, see Fig. \ref{fig:SNR_time_evolution_comparison}a,b. On the contrary in the 
$M_\star = 20\, \rm M_{\odot}$ model is much smoother, see studies on the close surroundings of runaway red supergiants~\citep{noriegacrespo_aj_114_1997,
2ssssssssssssssss,mohamed_aa_541_2012,meyer_mnras_506_2021}. 
\textcolor{black}{
Our model raises the question of the nature of the large-scale instabilities developing into the bow shock of the last pre-supernova stellar wind, as it is the first circumstellar structure that the blastwave will encounter when expanding past the snowplough phase. 
This is particularly pronounced in the supernova remnant generated by a $M_\star = 35\, \rm M_{\odot}$ progenitor star moving at velocity $v_{\star}=40\, \rm km\, \rm s^{-1}$, of potential numerical origin. This feature develops with our code for the combination of bulk motion and ISM number density that we consider and disappears in the model with lower $v_{\star}$, see Fig. \ref{fig:SNR_time_evolution_comparison}a,b, which witness a physical origin in the 
growth of such Rayleigh-Taylor-based instabilities~\citep{vishniac_apj_428_1994,vanmarle_aa_561_2014}, 
however, the role of the symmetry axis of our coordinate system in triggering the instabilities is known and should be kept in mind~\citep{MIGNONE2014784}. 
Full three-dimensional models of such systems would only permit to answer this question~\citep{blondin_na_57_1998,meyer_mnras_506_2021}, also exploring the stabilising 
role of the ambient medium magnetic field on the overall morphology of the supernova 
remnant and the mixing of material therein. 
}

At 
$8\, \rm kyr$, the supernova shock wave has reached and interacted with the bow shocks in each model (Fig. \ref{fig:SNR_time_evolution_comparison}c,d). 
Along the direction of stellar motion, one can see the effect of the mass that is trapped into the circumstellar medium, in the sense that the forward shock of the expanding blastwave has gone through the bow shock and propagates further into the unperturbed ISM as a mushroom-like outflow. This process is more important in the case of the 
supergiant progenitor than that of the $M_\star = 35\, \rm M_{\odot}$ 
progenitor~\citep{meyer_mnras_450_2015}. 
At times $80\, \rm kyr$, the reverse shock of the supernova is more advanced in its reverberation towards the centre of the explosion in the simulation with a $M_\star = 20\, \rm M_{\odot}$ progenitor, due to the small bow shock triggers it sooner. 
Nevertheless, the unstable character of this inward-moving shock front is more pronounced in the case of the model with a Wolf-Rayet supernova progenitor because the shock surface is wider. The instabilities had more time to grow due to the ejecta-wind strong contrast in density and velocity in the ejecta-circumstellar medium interface (green contours), see Fig. 
\ref{fig:SNR_time_evolution_comparison}e,f.

\subsection{Synchrotron emission maps}
\label{resultats_maps}

Fig. \ref{fig:sync_map_20_20kms} shows the $1.4\, \rm GHz$ emission maps of the supernova remnant generated by the moving progenitors with velocity $v_{\star}=20\, \rm km\, \rm s^{-1}$, considering viewing angles of $\theta_{\rm obs} = 0\degree$ (top panels) and $\theta_{\rm obs} = 45\degree$ (bottom panels), at $8\, \rm kyr$ (left panels) and $80\, \rm kyr$ (right panels) of evolution. These maps correspond to the case of a progenitor with $M_\star = 20\, \rm M_{\odot}$. 
The remnant in the model Run-20-MHD-20-SNR, for $\theta_{\rm obs} = 0\degree$, displays a barrel-like morphology at $8\, \rm kyr$ while a Cygnus-loop-like appearance  is observed at $80\, \rm kyr$ (Fig. \ref{fig:sync_map_20_20kms}a,b). If the remnant is observed with $\theta_{\rm obs} = 45\degree$, it takes the shape of rings (Fig. \ref{fig:sync_map_20_20kms}c,d). 
The model Run-35-MHD-20-SNR, involving a $M_\star = 35\, \rm M_{\odot}$ progenitor moving with velocity $v_{\star}=20\, \rm km\, \rm s^{-1}$, displays a horseshoe morphology if considered at times $8\, \rm kyr$ with an angle $\theta_{\rm obs} = 0\degree$ 
(Fig. \ref{fig:sync_map_35_20kms}a) that evolves to a bilateral morphology at $80\, \rm kyr$, see Fig. \ref{fig:sync_map_35_20kms}b. This remnant, seen under a viewing angle of $\theta_{\rm obs} = 45\degree$, has the shape of a ring and a bright arc.

Fig. \ref{fig:sync_map_20_40kms} displays the emission maps for the red supergiant model moving with the velocity $v_{\star}=40\, \rm km\, \rm s^{-1}$ before the explosion. Under a viewing angle of $\theta_{\rm obs} = 0\degree$, the remnant has a bilateral morphology that is closer to that of the Cygnus loop than in 
that with velocity $v_{\star}=20\, \rm km\, \rm s^{-1}$, which evolves to a rounder morphology at later time (Fig. \ref{fig:sync_map_20_40kms}a,b), while a 
viewing angle of $\theta_{\rm obs} = 45\degree$ produce a quasi-circular observed morphology in projection. 
The $M_\star = 35\, \rm M_{\odot}$ progenitor star moving with velocity 
$v_{\star}=40\, \rm km\, \rm s^{-1}$ displays an irregular morphology which originates 
from the more complex distribution of the circumstellar medium at the moment of the 
explosion (Fig. \ref{fig:sync_map_35_40kms}c,d). The supernova remnant appears as 
a large ovoidal structure reflecting the instabilities of the circumstellar medium, 
generated with the Wolf-Rayet stellar wind interacts with the previous wind bow shocks. 
In other words, no Cygnus-loop supernova remnants are produced when the 
fast-moving 
($v_{\star}=40\, \rm km\, \rm s^{-1}$) 
supernova progenitor undergoes an 
evolutionary phase beyond the red supergiant.

\begin{figure*}
        \centering
        \includegraphics[width=0.7\textwidth]{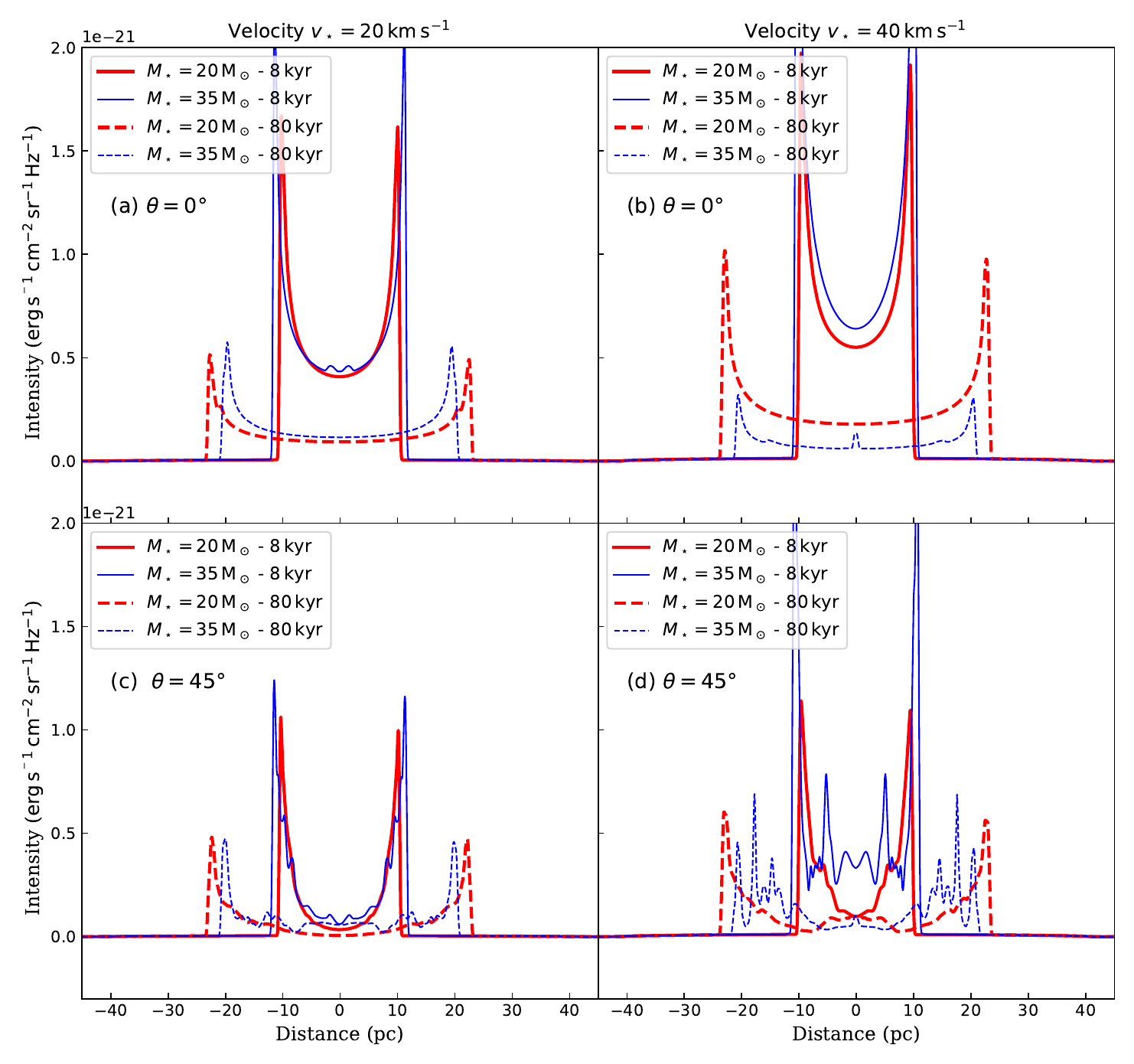}  \\    
        \includegraphics[width=0.7\textwidth]{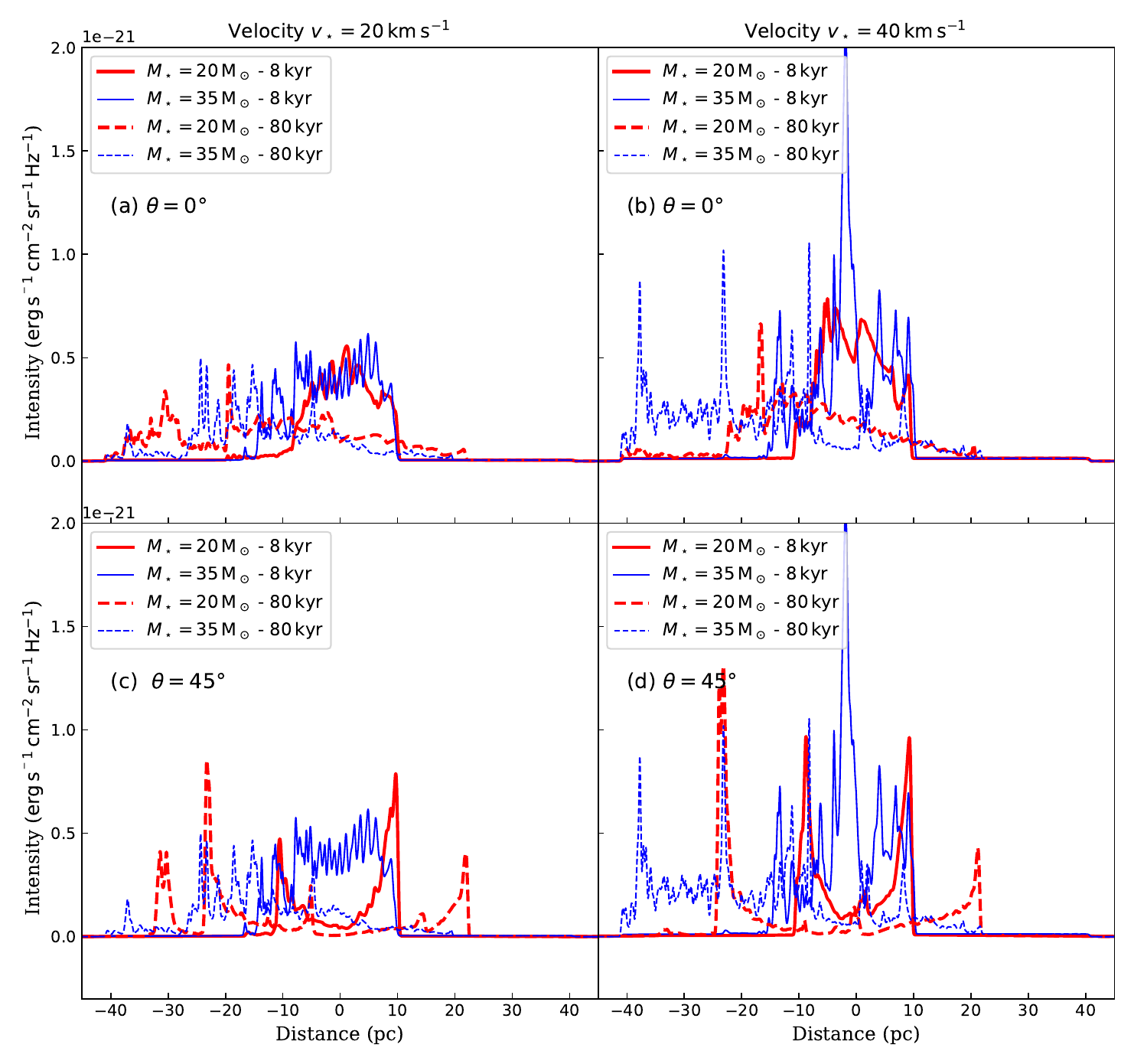}  \\          
        \caption{
        Horizontal (top) and vertical (bottom) cross-sections taken through 
        non-thermal $1.4\, \rm GHz$ synchrotron emission maps of the supernova 
        remnants.
        }
        \label{fig:cuts_horizontal}  
\end{figure*}

Fig.\ref{fig:cuts_horizontal} compares horizontal cross-sections taken through the emission 
maps of the supernova remnants, considered with viewing angle of $\theta_{\rm obs} = 0\degree$ 
(Fig. \ref{fig:cuts_horizontal}a,b), $\theta_{\rm obs} = 45\degree$ (
Fig.\ref{fig:cuts_horizontal}c,d), for the progenitors moving with velocities 
$v_{\star}=20\, \rm km\, \rm s^{-1}$ (Fig.\ref{fig:cuts_horizontal}a,c) and 
$v_{\star}=40\, \rm km\, \rm s^{-1}$ (Fig.\ref{fig:cuts_horizontal}b,d). 
Regardless of the viewing angle under which their remnants are considered for both progenitors, 
the synchrotron surface brightness diminishes with time, see solid and 
dashed lines at times $8\, \rm kyr$ and $80\, \rm kyr$ (Fig.\ref{fig:cuts_horizontal}a,c). 
The Wolf-Rayet remnant is slightly brighter than the red supergiant remnant, although it 
remains to the same order of magnitude in terms of surface brightnesses, which are 
$\approx 0.5-2.0 \times 10^{-21}\, \rm erg \rm s^{-1}\, \rm cm^{-2}\, \rm sr^{-1}\, \rm Hz^{-1}$. 
This increase in non-thermal emission is due to stronger shocks that form in these remnants, compressing the magnetic field better. 
%
The bottom panel of Fig.\ref{fig:cuts_horizontal} displays slices taken vertically through 
the supernova remnants. They are globally dimmer at $v_{\star}=20\, \rm km\, \rm s^{-1}$ 
and brighter at $v_{\star}=40\, \rm km\, \rm s^{-1}$ because of the denser filaments 
forming in them, for both viewing angles $\theta_{\rm obs} = 0\degree$ and 
$\theta_{\rm obs} = 45\degree$, respectively. 
%
Again, the differences arise later than the remnants' initial state 
(Fig.\ref{fig:cuts_horizontal}b,d). 
%
%

\begin{figure*}
        \centering
        \includegraphics[width=0.85\textwidth]{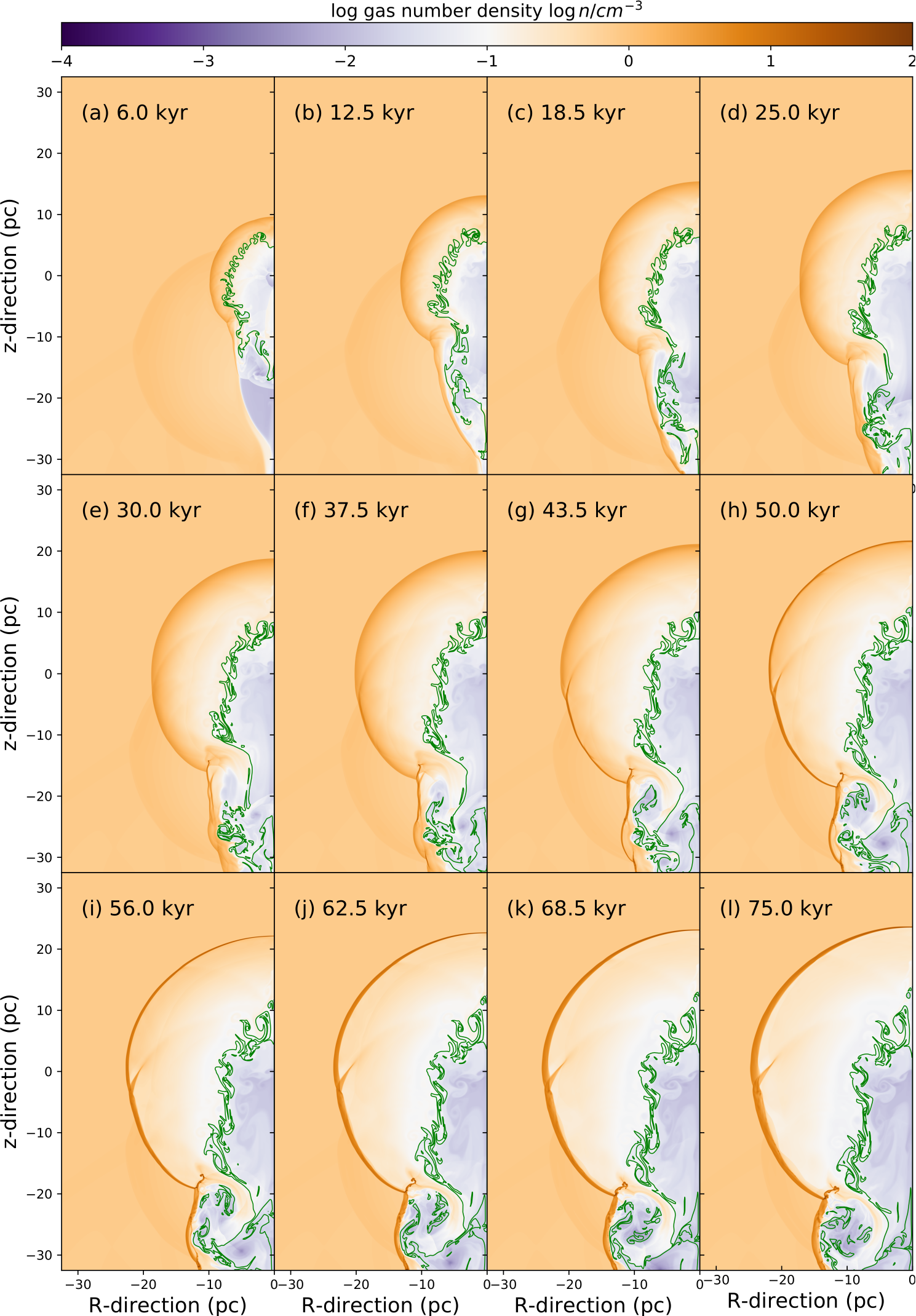}  \\
        \caption{
        \textcolor{black}{
        Time sequence evolution of the  number density fields in the model 
        corresponding to the most common supernova remnant, i.e. generated 
        by a $20\, \rm M_{\odot}$ moving with velocity 
        $v_{\star}=20\, \rm km\, \rm s^{-1}$. 
        The remnant morphology evolves from bilateral to Cygnus Loop. 
        The green contour mark the parts of the supernova remnants 
        with a $50\%$ contribution of ejecta. 
        }
        }
        \label{fig:SNR_time_evolution_comparison_details}  
\end{figure*}

\section{Discussion}
\label{discussion}

This section proposes a comparison of the results with real, available observations. 
We particularly concentrate our analysis on bilateral and Cygnus-loop-like supernova 
remnants. 

\subsection{Limitations of the model}
\label{future}

As in the previous papers of this series~\citep{meyer_mnras_521_2023}, the present 2.5-dimensional magnetohydrodynamic simulations would benefit from a full three-dimensional treatment of the wind-ISM interactions and that of the ejecta-bubble calculation. Such simulations will add realisticness to the solution to our problem, especially concerning the stability of the pre-supernova magnetised circumstellar medium  
\textcolor{black}{and fix the artificial ring-like artefacts arising from mapping 2.5-dimensional models into a 3-dimensional shape. It is not the maximum synchrotron surface brightness which would change, but rather the west-east symmetry concerning the $R=0$ direction that will 
disappear, as well as the series of horizontal bright lines ($\theta_{\rm obs}=0\degree$) 
and ring-like ($\theta_{\rm obs}>0\degree$) filaments located inside of the supernova 
remnant image, respectively. Those arcs will appear more ragged and clumpy while displaying more random distribution throughout the remnant's interior. An illustration of the 
effects of full 3-dimensional emission maps can be found within the example of the 
optical H$\alpha$ bow shock of runaway red supergiant stars in~\citet{meyer_2014bb} 
and ~\citet{meyer_mnras_506_2021}. 
}

The microphysical processes included in the numerical models also leave room for improvement, e.g. by including heat transfers or the ionisation of the supernova progenitor of the material of its circumstellar medium.
%
The multi-phased character of the Galactic plane of the Milky Way is also an ingredient in our numerical model, which would benefit from future improvements. Particularly, initiating the wind-ISM models in a medium including turbulence, granularity provided by colder clumps and hotter cavities could potentially affect the supernova blastwave's propagation and modify the current results. These issues will be considered in follow-up works. 
Last, the presence of a pulsar wind nebula could further modify the 
results~\citep{blondin_apj_563_2001,temim_apj_808_2015,temim_apj_851_2017,
temim_apj_932_2022}.

\subsection{A time-sequence evolution?}
\label{comp_obsrvations}

This work covers the parameter space of the most common runaway massive supernova progenitors to investigate their morphologies and non-thermal radio appearance. The models in our study reveal that most of them undergo the formation of a bilateral supernova remnant when their supernova blastwave propagates through the preshaped circumstellar medium. Indeed, all models but one, i.e. all remnants generated by a red supergiant progenitor ($M_\star \ge 20\, \rm M_{\odot}$) or 
by a Wolf-Rayet progenitor ($M_\star \ge 35\, \rm M_{\odot}$) that is a slow 
runaway ($v_\star \approx 20\, \rm km\, \rm s^{-1}$) induce a bilateral supernova 
remnant made of two arcs produced when the shock wave interacts with the walls of the low-density cavity left behind the progenitor when moving before the explosion, at least for the ISM densities we consider. 
%
Additionally, in the case of a red supergiant progenitor, the bilateral morphology always precedes that of a Cygnus-loop-like morphology. We propose that each Cygnus-loop supernova remnant first came through an earlier phase during which it harboured a bilateral morphology. However, a bilateral shape will mostly but not necessarily evolve.


\begin{figure}
        \centering
        \includegraphics[width=0.48\textwidth]{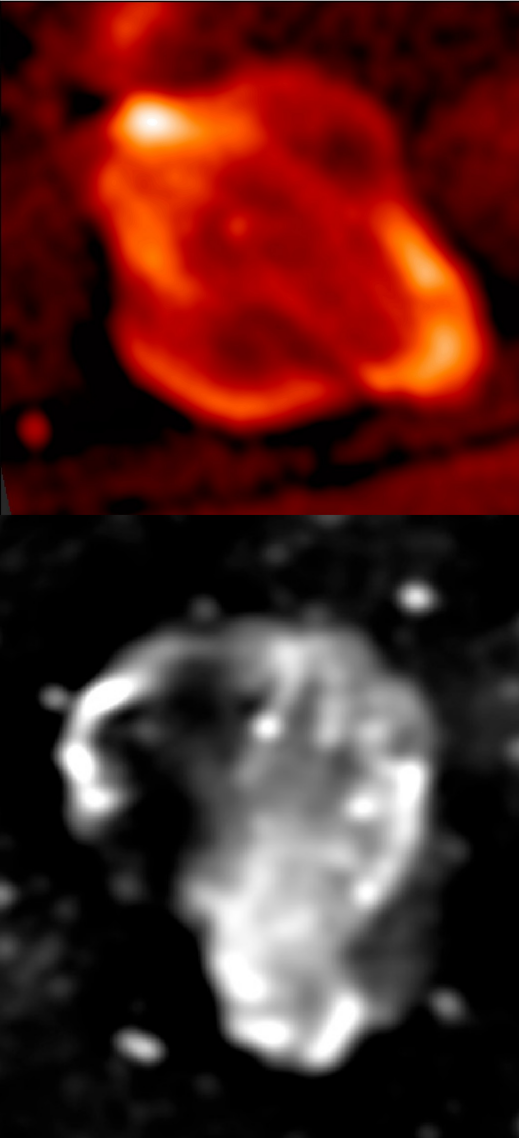}  \\             
        \caption{
        \textcolor{black}{
        Radio images show a bilateral G309.2-00.6~\citep[top 
        panel,][]{ferrand_adspr_49_2012} and the Cygnus 
        Loop~\citep[bottom panel,][]{uyaniker_426_aa_2004} 
        core-collapse supernova remnants. This study conjectures 
        that bilateral remnants from massive stars are less common 
        than those of Cygnus Loops but also that these shapes constitute 
        a time-sequence of morphological evolution affecting the 
        remnants of the most common runaway red supergiant progenitors.  
        }
        }
        \label{fig:obs_SNRs}  
\end{figure}

In particular, we specifically conjecture that the Cygnus-loop nebula had once been in a bilateral shape. We also propose that the same must apply to Simeis 147, a supernova remnant of similar observed features: a bulb shape appears by the interaction between the blastwave and the circumstellar medium of its progenitor.

\subsection{The occurrence of Cygnus-loop-like supernova remnants}
\label{future}

The initial mass of massive stars as they form in the cold phase of the ISM 
of the Milky Way is determined by the so-called initial mass function, 
which, for high-mass stellar objects, reads, 
\begin{equation}
   dN( M_{\star} ) \propto M_{\star}^{-\zeta}, 
   \label{eq:imf1}
\end{equation}
where $dN$ is the differential element of massive stars and $\zeta=2.3$ \citep[see~][]{kroupa_mnras_322_2001}. Similarly, the space velocities of 
runaway massive stars as produced when a component of a massive binary 
system explodes is, 
\begin{equation}
   dN(v_\star) \propto e^{ -v_\star / v_{\rm max} }, 
   \label{eq:vel1}
\end{equation}
with $v_{\rm max}=150\, \rm km\, \rm s^{-1}$ \citep[see][]{bromley_apj_706_2009}.  
\textcolor{black}{
The above mentioned distributions indicate that (i) most core-collapse 
supernova progenitors are massive stars in the lower region of the high 
mass spectrum (typically ending their lives as red supergiants and 
(ii) that runaway massive stars mostly move with small Mach numbers 
through the ISM (with typical supersonic velocities $v_\star \approx 
20\, \rm km\, \rm s^{-1}$ in the plane of the Galaxy). 
}

Our sample of supernova progenitors and space velocities (see Table \ref{tab:table1}) 
therefore covers the parameter space of the most common runaway progenitors and half of them, those 
involving a red supergiant progenitor, generate a Cygnus-loop morphology during 
the older times of their evolution. Since such progenitors are the most common 
runaway high-mass objects, we conjecture that a majority of supernova remnants of 
moving objects, i.e. that are either in the field or in the high latitudes of the
Milky Way, display, or are in the process of forming a Cygnus-loop-like remnant. 
These objects are fainter than those of higher-mass progenitors, and, therefore, 
more difficult to observe. Consequently the supernova remnants to be discovered 
in the next decade by means of, e.g. the current {\it James Webb Space Telescope} 
(JWST) or the forthcoming {\it Cherenkov Telescope Array} (CTA) observatories, 
see ~\citep{2023MNRAS.523.5353A,2023APh...15002850A,2023arXiv230903712C}, will, 
according to our results, 
\textcolor{black}{
reveal a majority of supernova remnants that are either in the bilateral or in the 
Cygnus-loop case, with a subset of the sooner gong to further evolve to the later.  
}

\subsection{Predictions for the most common supernova remnants}
\label{prediction}

\textcolor{black}{
Fig. \ref{fig:SNR_time_evolution_comparison_details} displays the number density field 
in the simulation of the supernova remnant of initial mass $20\, \rm M_{\odot}$ 
moving with velocity $v_{\star}=20\, \rm km\, \rm s^{-1}$ through the ISM, 
for times spanning from $6\, \rm kyr$ to $75\, \rm kyr$. The green line marks 
the location of the discontinuity between supernova ejecta and the other kind 
of material (stellar wind and ISM). 
At early times the blastwave interacts strongly with the lateral region of the 
circumstellar medium, i.e. the wings of the stellar wind bow shock, which 
generates a bilateral shape, see Fig. \ref{fig:SNR_time_evolution_comparison_details}a-d.  
The Cygnus Loop morphology begins to form at time $25\, \rm kyr$ after the 
onset of the explosion, with its characteristic shape made of two components: 
the overall mushroom of expanding supernova ejecta plus the tube of channelled 
material into the low-density cavity produced behind the progenitor as a result 
of its bluk motion through the ambient medium. 
The density of the shocked layer of ejecta, ISM and stellar wind that is 
swept-up by the forward shock of the supernova blastwave increases with time, 
compressing the local magnetic field better and enhancing the synchrtron 
emissivity (Fig. \ref{fig:SNR_time_evolution_comparison_details}j-l). 
We conclude that this remnant spends about $18.5/80.0 \approx 23 \%$ of 
its early evolution time in the bilateral shape, while it harbors a Cygnus 
Loop morphology the rest of the time. Our results indicate therefore that 
the population of Cygnus Loop is $3$ times superior to that of bilateral 
remnants, at least for the conditions of ambient medium that we consider. 
However, the sooner are brighter than the later (Fig. \ref{fig:sync_map_20_20kms}), 
which is consistent with the more numerous detection of bilateral objects than 
Cygnus Loop amongst the known population of middle-aged to older core-collapse 
supernova remnants. Future high-resolution telescopes will help in building 
larger population statistics, which will better tests our prediction in the 
future. 
}

\textcolor{black}{
How close are our predictions to real data? We plot in Fig. \ref{fig:obs_SNRs} 
the radio appearance of two core-collapse supernova remnants of bilateral 
and Cygnus Loop morphology, respectively. G309.2-00.6 is a bilateral remnant of 
age $\le\, \rm 4\, \rm kyr$ and size $2$$-$$7\, \rm kpc$ that has a complex shape, 
superimposing large symmetric equatorial ears with two polar arcs, as seen in 
the top panel of Fig. \ref{fig:obs_SNRs}. Literature on G309.2-00.6 is mostly 
observational. It soon revealed strong deviations of its morphology from the 
Sedov-Taylor solution, requiring the inclusion of circumstellar material pre-shaped 
by the progenitor’s stellar winds~\citep{gvaramadez_1999} and an explosive jet 
responsible for the second series of protuberances~\citep{gaensler_mnras_299_1998,
gaensler_phd_1999}. Its size and age are consistent with our model while the model 
with $20\, \rm M_{\odot}$ and $v_{\star}=20\, \rm km\, \rm s^{-1}$, see 
Fig. \ref{fig:SNR_time_evolution_comparison_details}a,b. Differences in the shape might 
be caused by the explosive jet that we do not model~\citep{soker_raa_2023} and/or 
by a different viewing angle of the supernova remnant. 
The Cygnus Loop is a peculiar supernova remnant that 
has been more studied, both by means of observations and with numerical models, 
than G309.2-00.6. Initially interpreted as a champagne outflow produced by a 
supernova explosion located at the edge of a molecular cloud~\citep{aschenbach_aa_341_1999}, 
the Cygnus Loop nebula might be the outflow of a blast wave from a stellar wind bow 
shock~\citep{meyer_mnras_450_2015,fang_mnras_464_2017}. Its size is estimated 
to be about $37\, \rm pc$ and its age $\approx 20\, \rm kyr$, which is in 
accordance with our same numerical model, see Fig. 
\ref{fig:SNR_time_evolution_comparison_details}c,d. Hence, both remnants 
of different typical morphologies can qualitatively be explained by our model 
with $20\, \rm M_{\odot}$ and $v_{\star}=20\, \rm km\, \rm s^{-1}$.
}


\section{Conclusion}
\label{conclusion}

In this study, we investigate the differences between the non-thermal synchrotron 
emission of supernova remnants generated by moving massive stars. Our parameter space 
covers the most common of such core-collapse supernova progenitors, of zero-age main-sequence 
mass $M_\star = 20$ and $M_\star = 35\, \rm M_{\odot}$, ending their lives as a red 
supergiant or as a Wolf-Rayet star, and moving into the warm phase of the Galactic 
plane with bulk velocities $v_\star=20\, \rm km\, \rm s^{-1}$ and 
$v_\star=40\, \rm km\, \rm s^{-1}$, respectively. The evolution of the supernova remnants is followed from the onset of the explosion to older ages ($80\, \rm kyr$). 
The methodology consists of performing 2.5-dimensional magneto-hydrodynamics numerical 
simulations of supernova blastwaves released into the circumstellar medium shaped by the 
interaction of the stellar wind of these moving progenitors with the ambient medium, 
see~\citet{meyer_mnras_521_2023}. 
Synchrotron emission calculations complete the magnetohydrodynamical structure to generate radio maps with physical units, instead of the usual normalised maps presented 
in most papers, that we compare to observations.

We find that most supernova remnants generated by a runaway massive star should undergo a phase exhibiting a bilateral morphology but that only those either generated by a red supergiant or, to a lesser extent, produced by a slightly supersonically moving Wolf-Rayet evolving progenitor, further develop as a Cygnus-loop-like supernova remnant~\citep{aschenbach_aa_341_1999,meyer_mnras_450_2015,fang_mnras_464_2017}. 
Our $1.4\, \rm GHz$ non-thermal emission maps~\citep{velazquez_mnras_519_2023,
2023arXiv230916410V} indicate that supernova remnants reveal duller projected intensities as the blastwave propagates through the defunct stellar surroundings 
and the unperturbed ISM and the remnant become old. 
In all explored scenarios, supernova remnants generated by faster-moving stars are brighter than those induced by stars, which were moving slower before dying. 
Their Cygnus-loop appearances do not survive when the viewing angle between the 
observer and the place of the sky is important ($\theta_{\rm obs} = 45\degree$).

\textcolor{black}{ 
Our methodology permits to investigate the time-dependant evolution of 
supernova remnants' brightness, and, in future studies, we will apply it 
in the context of (plerionic) supernova remnants of static massive stars 
\citep{mnras_mnras_515_2022,meyer_mnras_527_2024}. 
}
Our work particularly proposes a non-thermal synchrotron 
bilateral to Cygnus-loop-like morphological time sequence evolution 
for the appearance of Galactic supernova 
remnants, and that it should affect most of the dead stellar 
surroundings in the field since red supergiant 
progenitors are more common than heavier exploding stars. 
This implies that among the many supernova remnants to be discovered in the 
future, e.g. by means of the forthcoming {\it Cherenkov Telescope Array} (CTA),  
observatory~\citep{2023MNRAS.523.5353A,2023APh...15002850A,2023arXiv230903712C}, 
a significant fraction of them should be undergoing this bilateral-to-Cygnus-loop 
evolutionary sequence.


\begin{acknowledgements}
The authors acknowledge the North-German Supercomputing Alliance (HLRN) for providing HPC 
resources that have contributed to the research results reported in this paper. 
M.~Petrov acknowledges the Max Planck Computing and Data Facility (MPCDF) for providing data 
storage and HPC resources, which contributed to testing and optimising the {\sc pluto} code. PFV acknowledge financial support from PAPIIT-UNAM grant IG100422. MV is a doctoral fellow of CONICET, Argentina.

This work has been supported by the grant PID2021-124581OB-I00 funded by 
MCIU/AEI/10.13039/501100011033 and 2021SGR00426 of the Generalitat de Catalunya. 
This work was also supported by the Spanish program Unidad de Excelencia Mar\' ia 
de Maeztu CEX2020-001058-M.
This work also supported with funding from the European Union 
NextGeneration program (PRTR-C17.I1). 
\end{acknowledgements}


\bibliographystyle{aa} 
\bibliography{grid} 

\begin{thebibliography}{140}
\expandafter\ifx\csname natexlab\endcsname\relax\def\natexlab#1{#1}\fi

\bibitem[{{Abbott}(1980)}]{1980ApJ...242.1183A}
{Abbott}, D.~C. 1980, \apj, 242, 1183

\bibitem[{{Abbott}(1982)}]{1982ApJ...259..282A}
{Abbott}, D.~C. 1982, \apj, 259, 282

\bibitem[{{Acero} {et~al.}(2023){Acero}, {Acharyya}, {Adam}, {Aguasca-Cabot},
  {Agudo}, {Aguirre-Santaella}, {Alfaro}, {Aloisio}, {Crespo}, {Batista},
  {Amati}, {Amato}, {Ambrosi}, {Ang{\"u}ner}, {Aramo}, {Arcaro}, {Armstrong},
  {Asano}, {Ascasibar}, {Aschersleben}, {Backes}, {Baktash}, {Balazs}, {Balbo},
  {Ballet}, {Larriva}, {Martins}, {de Almeida}, {Barrio}, {Bastieri}, {Baxter},
  {Tjus}, {Benbow}, {Bernardos-Mart{\'\i}n}, {Bernete}, {Berti}, {Bertucci},
  {Beshley}, {Bhattacharjee}, {Bhattacharyya}, {Biland}, {Bissaldi}, {Biteau},
  {Blanch}, {Bordas}, {Bottacini}, {Bregeon}, {Brose}, {Bucciantini},
  {Bulgarelli}, {Capasso}, {Dolcetta}, {Caraveo}, {Cardillo}, {Carosi},
  {Casanova}, {Cascone}, {Cassol}, {Catalani}, {Cerruti}, {Chadwick}, {Chaty},
  {Chen}, {Chernyakova}, {Chiavassa}, {Chudoba}, {Coimbra-Araujo}, {Conforti},
  {Contreras}, {Costa}, {Costantini}, {Cristofari}, {Crocker}, {D'Amico},
  {D'Ammando}, {De Angelis}, {De Caprio}, {de Gouveia Dal Pino}, {de Ona
  Wilhelmi}, {de Souza}, {Delgado}, {della Volpe}, {Depaoli}, {Di Girolamo},
  {Di Pierro}, {Di Tria}, {Di Venere}, {Diebold}, {Djuvsland}, {Donini},
  {Doro}, {Dos Anjos}, {Dwarkadas}, {Einecke}, {Els{\"a}sser}, {Emery},
  {Evoli}, {Falceta-Goncalves}, {Fedorova}, {Fegan}, {Ferrand}, {Fiandrini},
  {Filipovic}, {Fioretti}, {Fiori}, {Foffano}, {Fontaine}, {Fukami}, {Galanti},
  {Galaz}, {Gammaldi}, {Gasbarra}, {Ghalumyan}, {Ghirlanda}, {Giarrusso},
  {Giavitto}, {Giglietto}, {Giordano}, {Giroletti}, {Giuliani}, {Giunti},
  {Godinovic}, {Coelho}, {Gr{\'e}aux}, {Green}, {Grondin}, {Gueta}, {Gunji},
  {Hassan}, {Heller}, {Hern{\'a}ndez-Cadena}, {Hinton}, {Hnatyk}, {Hnatyk},
  {Hoffmann}, {Hofmann}, {Holder}, {Horan}, {Horvath}, {Hrabovsky}, {Hrupec},
  {Inada}, {Incardona}, {Inoue}, {Ishio}, {Jamrozy}, {Janecek},
  {Mart{\'\i}nez}, {Jin}, {Jung-Richardt}, {Jurysek}, {Kaaret}, {Karas},
  {Katz}, {Kerszberg}, {Kh{\'e}lifi}, {Kieda}, {Kissmann}, {Kleiner}, {Kluge},
  {Kluzniak}, {Kn{\"o}dlseder}, {Kobayashi}, {Kohri}, {Komin}, {Kornecki},
  {Kubo}, {La Palombara}, {L{\'a}inez}, {Lamastra}, {Lapington},
  {Lemoine-Goumard}, {Lenain}, {Leone}, {Leto}, {Leuschner}, {Lindfors},
  {Liodakis}, {Lohse}, {Lombardi}, {Longo}, {L{\'o}pez-Coto}, {L{\'o}pez-Moya},
  {L{\'o}pez-Oramas}, {Loporchio}, {Luque-Escamilla}, {Macias}, {Mackey},
  {Majumdar}, {Mandat}, {Manganaro}, {Manic{\`o}}, {Marconi}, {Mart{\'\i}},
  {Mart{\'\i}nez}, {Martinez}, {Martinez}, {Mello}, {Menchiari}, {Meyer},
  {Micanovic}, {Miceli}, {Miceli}, {Michalowski}, {Miener}, {Miranda},
  {Mitchell}, {Mode}, {Moderski}, {Mohrmann}, {Molina}, {Montaruli},
  {Morcuende}, {Morlino}, {Morselli}, {Mos{\`e}}, {Moulin}, {Mukherjee},
  {Munari}, {Murach}, {Nagai}, {Nagataki}, {Nemmen}, {Niemiec}, {Nieto},
  {Rosillo}, {Nikolajuk}, {Nishijima}, {Noda}, {Novosyadlyj}, {Nozaki},
  {Ohishi}, {Ohm}, {Ohtani}, {Okumura}, {Olmi}, {Ong}, {Orienti}, {Orito},
  {Orlandini}, {Orlando}, {Orlando}, {Ostrowski}, {Oya}, {Pantaleo}, {Paredes},
  {Patricelli}, {Pecimotika}, {Peresano}, {P{\'e}rez-Romero}, {Persic},
  {Petruk}, {Piano}, {Pietropaolo}, {Pirola}, {Pittori}, {Pohl}, {Ponti},
  {Prandini}, {Principe}, {Priyadarshi}, {Pueschel}, {P{\"u}hlhofer}, {Pumo},
  {Quirrenbach}, {Rando}, {Razzaque}, {Reichherzer}, {Reimer}, {Reimer},
  {Renaud}, {Reposeur}, {Rib{\'o}}, {Richtler}, {Rico}, {Rieger}, {Rigoselli},
  {Riitano}, {Rizi}, {Roache}, {Romano}, {Romeo}, {Rosado}, {Rowell}, {Rudak},
  {Sadeh}, {Safi-Harb}, {Saha}, {Sailer}, {S{\'a}nchez-Conde}, {Sarkar},
  {Satalecka}, {Saturni}, {Scherer}, {Schov{\'a}nek}, {Schussler}, {Schwanke},
  {Scuderi}, {Seglar-Arroyo}, {Sergijenko}, {Servillat}, {Shang}, {Sharma},
  {Siejkowski}, {Sliusar}, {S{\l}owikowska}, {Sol}, {Specovius}, {Spencer},
  {Spengler}, {Stamerra}, {Stani{\v{c}}}, {Starecki}, {Starling}, {Stolarczyk},
  {Pereira}, {Suda}, {Suomijarvi}, {Sushch}, {Tajima}, {Tam}, {Tanaka},
  {Tavecchio}, {Testa}, {Tian}, {Tibaldo}, {Torres}, {Tothill}, {Vallage},
  {Vallania}, {van Eldik}, {van Scherpenberg}, {Vandenbroucke}, {Acosta},
  {Vecchi}, {Vercellone}, {Verna}, {Viana}, {Vignatti}, {Vitale}, {Vodeb},
  {Vorobiov}, {Vuillaume}, {Wagner}, {Walter}, {White}, {Wierzcholska}, {Will},
  {Williams}, {Yang}, {Yoshida}, {Yoshikoshi}, {Zaharijas}, {Zampieri},
  {Zavrtanik}, {Zavrtanik}, {Zhdanov}, {Z̆ivec}, \& {Cherenkov Telescope Array
  Consortium}}]{2023APh...15002850A}
{Acero}, F., {Acharyya}, A., {Adam}, R., {et~al.} 2023, Astroparticle Physics,
  150, 102850

\bibitem[{{Acharyya} {et~al.}(2023){Acharyya}, {Adam}, {Aguasca-Cabot},
  {Agudo}, {Aguirre-Santaella}, {Alfaro}, {Aloisio}, {Alves Batista}, {Amato},
  {Ang{\"u}ner}, {Aramo}, {Arcaro}, {Asano}, {Aschersleben}, {Ashkar},
  {Backes}, {Baktash}, {Balazs}, {Balbo}, {Ballet}, {Bamba}, {Baquero Larriva},
  {Barbosa Martins}, {Barres de Almeida}, {Barrio}, {Bastieri}, {Batista},
  {Batkovic}, {Baxter}, {Becerra Gonz{\'a}lez}, {Becker Tjus}, {Benbow},
  {Bernardini}, {Bernardos Mart{\'\i}n}, {Bernete Medrano}, {Berti},
  {Bertucci}, {Beshley}, {Bhattacharjee}, {Bhattacharyya}, {Bigongiari},
  {Biland}, {Bissaldi}, {Bocchino}, {Bordas}, {Borkowski}, {Bottacini},
  {B{\"o}ttcher}, {Bradascio}, {Brown}, {Bulgarelli}, {Burmistrov}, {Caroff},
  {Carosi}, {Carqu{\'\i}n}, {Casanova}, {Cascone}, {Cassol}, {Cerruti},
  {Chadwick}, {Chaty}, {Chen}, {Chiavassa}, {Chytka}, {Conforti}, {Cortina},
  {Costa}, {Costantini}, {Cotter}, {Crestan}, {Cristofari}, {D'Ammando},
  {Dalchenko}, {Dazzi}, {De Angelis}, {De Caprio}, {de Gouveia Dal Pino}, {De
  Martino}, {de Naurois}, {de Souza}, {del Valle}, {Delgado Giler}, {Delgado},
  {della Volpe}, {Depaoli}, {Di Girolamo}, {Di Piano}, {Di Pierro}, {Di Tria},
  {Di Venere}, {Diebold}, {Doro}, {Dumora}, {Dwarkadas}, {Eckner}, {Egberts},
  {Emery}, {Escudero}, {Falceta-Goncalves}, {Fedorova}, {Fegan}, {Feng},
  {Ferenc}, {Ferrand}, {Fiandrini}, {Filipovic}, {Fioretti}, {Foffano},
  {Fontaine}, {Fukui}, {Gaggero}, {Galanti}, {Galaz}, {Gallozzi}, {Gammaldi},
  {Garczarczyk}, {Gasbarra}, {Gasparrini}, {Ghalumyan}, {Giarrusso},
  {Giavitto}, {Giglietto}, {Giordano}, {Giuliani}, {Glicenstein}, {Goldoni},
  {Goulart Coelho}, {Granot}, {Green}, {Green}, {Grondin}, {Gueta}, {Hadasch},
  {Hamal}, {Hassan}, {Hayashi}, {Heller}, {Hern{\'a}ndez Cadena}, {Hiroshima},
  {Hnatyk}, {Hnatyk}, {Hofmann}, {Holder}, {Holler}, {Horan}, {Horvath},
  {Hrabovsky}, {H{\"u}tten}, {Iarlori}, {Inada}, {Incardona}, {Inoue}, {Iocco},
  {Jamrozy}, {Jin}, {Jung-Richardt}, {Jury{\v{s}}ek}, {Kantzas}, {Karas},
  {Katagiri}, {Kerszberg}, {Kn{\"o}dlseder}, {Komin}, {Kornecki}, {Kosack},
  {Kowal}, {Kubo}, {Lamastra}, {Lapington}, {Lemoine-Goumard}, {Lenain},
  {Leone}, {Leto}, {Leuschner}, {Lindfors}, {Lohse}, {Lombardi}, {Longo},
  {L{\'o}pez-Coto}, {L{\'o}pez-Oramas}, {Loporchio}, {Luque-Escamilla},
  {Macias}, {Majumdar}, {Mandat}, {Mangano}, {Manic{\`o}}, {Mariotti},
  {Marquez}, {Marsella}, {Mart{\'\i}}, {Martin}, {Mart{\'\i}nez}, {Mazin},
  {Menchiari}, {Meyer}, {Miceli}, {Miceli}, {Micha{\l}owski}, {Mitchell},
  {Moderski}, {Mohrmann}, {Molero}, {Molina}, {Montaruli}, {Moralejo},
  {Morcuende}, {Morselli}, {Moulin}, {Moya}, {Mukherjee}, {Munari},
  {Muraczewski}, {Nagataki}, {Nakamori}, {Nayak}, {Niemiec}, {Nievas},
  {Niko{\l}ajuk}, {Nishijima}, {Noda}, {Nosek}, {Novosyadlyj}, {Nozaki},
  {Ohishi}, {Ohm}, {Okumura}, {Olmi}, {Ong}, {Orienti}, {Orito}, {Orlandini},
  {Orlando}, {Orlando}, {Ostrowski}, {Oya}, {Pagliaro}, {Palatka}, {Pantaleo},
  {Paoletti}, {Paredes}, {Parmiggiani}, {Patricelli}, {Pech}, {Pecimotika},
  {Persic}, {Petruk}, {Pierre}, {Pietropaolo}, {Pirola}, {Pohl}, {Prandini},
  {Priyadarshi}, {P{\"u}hlhofer}, {Pumo}, {Punch}, {Queiroz}, {Quirrenbach},
  {Rain{\`o}}, {Rando}, {Razzaque}, {Reimer}, {Reimer}, {Reposeur}, {Rib{\'o}},
  {Richtler}, {Rico}, {Rieger}, {Rigoselli}, {Rizi}, {Roache}, {Rodriguez
  Fernandez}, {Romano}, {Romeo}, {Rosado}, {Rosales de Leon}, {Rudak},
  {Rulten}, {Sadeh}, {Saito}, {S{\'a}nchez-Conde}, {Sano}, {Santangelo},
  {Santos-Lima}, {Sarkar}, {Saturni}, {Scherer}, {Schovanek}, {Schussler},
  {Schwanke}, {Sergijenko}, {Servillat}, {Siejkowski}, {Siqueira}, {Spencer},
  {Stamerra}, {Stani{\v{c}}}, {Steppa}, {Stolarczyk}, {Suda}, {Tavernier},
  {Teshima}, {Tibaldo}, {Torres}, {Tothill}, {Vacula}, {Vallage}, {Vallania},
  {van Eldik}, {V{\'a}zquez Acosta}, {Vecchi}, {Ventura}, {Vercellone},
  {Viana}, {Vigorito}, {Vink}, {Vitale}, {Vodeb}, {Vorobiov}, {Vuillaume},
  {Wagner}, {Walter}, {White}, {Wierzcholska}, {Will}, {Yamazaki}, {Yang},
  {Yoshikoshi}, {Zacharias}, {Zaharijas}, {Zavrtanik}, {Zavrtanik},
  {Zdziarski}, {Zhdanov}, {Zi{\k{e}}tara}, \&
  {{\v{Z}}ivec}}]{2023MNRAS.523.5353A}
{Acharyya}, A., {Adam}, R., {Aguasca-Cabot}, A., {et~al.} 2023, \mnras, 523,
  5353

\bibitem[{{Aschenbach} \& {Leahy}(1999)}]{aschenbach_aa_341_1999}
{Aschenbach}, B. \& {Leahy}, D.~A. 1999, \aap, 341, 602

\bibitem[{{Asplund} {et~al.}(2009){Asplund}, {Grevesse}, {Sauval}, \&
  {Scott}}]{asplund_araa_47_2009}
{Asplund}, M., {Grevesse}, N., {Sauval}, A.~J., \& {Scott}, P. 2009, \araa, 47,
  481

\bibitem[{{Baalmann} {et~al.}(2020){Baalmann}, {Scherer}, {Fichtner},
  {Kleimann}, {Bomans}, \& {Weis}}]{baalmann_aa_634_2020}
{Baalmann}, L.~R., {Scherer}, K., {Fichtner}, H., {et~al.} 2020, \aap, 634, A67

\bibitem[{{Baalmann} {et~al.}(2021){Baalmann}, {Scherer}, {Kleimann},
  {Fichtner}, {Bomans}, \& {Weis}}]{baalmann_aa_650_2021}
{Baalmann}, L.~R., {Scherer}, K., {Kleimann}, J., {et~al.} 2021, \aap, 650, A36

\bibitem[{{Bear} \& {Soker}(2017)}]{bear_mnras_468_2017}
{Bear}, E. \& {Soker}, N. 2017, \mnras, 468, 140

\bibitem[{{Bear} \& {Soker}(2018)}]{2018MNRAS.478..682B}
{Bear}, E. \& {Soker}, N. 2018, \mnras, 478, 682

\bibitem[{{Bear} \& {Soker}(2021)}]{bear_mnras_500_2021}
{Bear}, E. \& {Soker}, N. 2021, \mnras, 500, 2850

\bibitem[{{Blondin} {et~al.}(2001){Blondin}, {Chevalier}, \&
  {Frierson}}]{blondin_apj_563_2001}
{Blondin}, J.~M., {Chevalier}, R.~A., \& {Frierson}, D.~M. 2001, \apj, 563, 806

\bibitem[{{Blondin} \& {Koerwer}(1998)}]{blondin_na_57_1998}
{Blondin}, J.~M. \& {Koerwer}, J.~F. 1998, \na, 3, 571

\bibitem[{{Brighenti} \&
  {D'Ercole}(1995{\natexlab{a}})}]{brighenti_mnras_277_1995}
{Brighenti}, F. \& {D'Ercole}, A. 1995{\natexlab{a}}, \mnras, 277, 53

\bibitem[{{Brighenti} \&
  {D'Ercole}(1995{\natexlab{b}})}]{brighenti_mnras_273_1995}
{Brighenti}, F. \& {D'Ercole}, A. 1995{\natexlab{b}}, \mnras, 273, 443

\bibitem[{{Broersen} {et~al.}(2014){Broersen}, {Chiotellis}, {Vink}, \&
  {Bamba}}]{broersen_mnras_441_2014}
{Broersen}, S., {Chiotellis}, A., {Vink}, J., \& {Bamba}, A. 2014, \mnras, 441,
  3040

\bibitem[{{Bromley} {et~al.}(2009){Bromley}, {Kenyon}, {Brown}, \&
  {Geller}}]{bromley_apj_706_2009}
{Bromley}, B.~C., {Kenyon}, S.~J., {Brown}, W.~R., \& {Geller}, M.~J. 2009,
  \apj, 706, 925

\bibitem[{{Brott} {et~al.}(2011{\natexlab{a}}){Brott}, {de Mink}, {Cantiello},
  {Langer}, {de Koter}, {Evans}, {Hunter}, {Trundle}, \&
  {Vink}}]{brott_aa_530_2011a}
{Brott}, I., {de Mink}, S.~E., {Cantiello}, M., {et~al.} 2011{\natexlab{a}},
  \aap, 530, A115

\bibitem[{{Brott} {et~al.}(2011{\natexlab{b}}){Brott}, {Evans}, {Hunter}, {de
  Koter}, {Langer}, {Dufton}, {Cantiello}, {Trundle}, {Lennon}, {de Mink},
  {Yoon}, \& {Anders}}]{brott_aa_530_2011b}
{Brott}, I., {Evans}, C.~J., {Hunter}, I., {et~al.} 2011{\natexlab{b}}, \aap,
  530, A116

\bibitem[{{Castro} {et~al.}(2017){Castro}, {Fossati}, {Hubrig}, {J{\"a}rvinen},
  {Przybilla}, {Nieva}, {Ilyin}, {Carroll}, {Sch{\"o}ller}, {Langer},
  {Schneider}, {Sim{\'o}n-D{\'\i}az}, {Morel}, {Butler}, \& {BOB
  Collaboration}}]{castro_aa_597_2017}
{Castro}, N., {Fossati}, L., {Hubrig}, S., {et~al.} 2017, \aap, 597, L6

\bibitem[{{Castro} {et~al.}(2015){Castro}, {Fossati}, {Hubrig},
  {Sim{\'o}n-D{\'\i}az}, {Sch{\"o}ller}, {Ilyin}, {Carrol}, {Langer}, {Morel},
  {Schneider}, {Przybilla}, {Herrero}, {de Koter}, {Oskinova}, {Reisenegger},
  {Sana}, \& {BOB Collaboration}}]{castro_aa_581_2015}
{Castro}, N., {Fossati}, L., {Hubrig}, S., {et~al.} 2015, \aap, 581, A81

\bibitem[{{Cherenkov Telescope Array Consortium} {et~al.}(2023){Cherenkov
  Telescope Array Consortium}, {:}, {Abe}, {Abe}, {Acero}, {Acharyya}, {Adam},
  {Aguasca-Cabot}, {Agudo}, {Aguirre-Santaella}, {Alfaro}, {Alfaro},
  {Alvarez-Crespo}, {Alves Batista}, {Amans}, {Amato}, {Ang{\"u}ner},
  {Antonelli}, {Aramo}, {Araya}, {Arcaro}, {Arrabito}, {Asano},
  {Ascas{\'\i}bar}, {Aschersleben}, {Ashkar}, {Stuani}, {Baack}, {Backes},
  {Baktash}, {Balazs}, {Balbo}, {Ballester}, {Baquero Larriva}, {Barbosa
  Martins}, {Barres de Almeida}, {Barrio}, {Batista}, {Batkovic}, {Batzofin},
  {Baxter}, {Becerra Gonz{\'a}lez}, {Beck}, {Becker Tjus}, {Benbow}, {Bernete
  Medrano}, {Bernl{\"o}hr}, {Berti}, {Bertucci}, {Beshley}, {Bhattacharjee},
  {Bhattacharyya}, {Bi}, {Biederbeck}, {Biland}, {Bissaldi}, {Biteau},
  {Blanch}, {Blazek}, {Boisson}, {Bolmont}, {Bordas}, {Bosnjak}, {Bottacini},
  {Bradascio}, {Braiding}, {Bronzini}, {Brose}, {Brown}, {Brun}, {Brunetti},
  {Bucciantini}, {Bulgarelli}, {Burelli}, {Burmistrov}, {Burton}, {Bylund},
  {Calisse}, {Campoy-Ordaz}, {Cantlay}, {Capalbi}, {Caproni},
  {Capuzzo-Dolcetta}, {Caraveo}, {Caroff}, {Carosi}, {Carquin}, {Carrasco},
  {Cascone}, {Cassol}, {Castro-Tirado}, {Cerasole}, {Cerruti}, {Chadwick},
  {Chaty}, {Chen}, {Chernyakova}, {Chiavassa}, {Chudoba}, {Chytka},
  {Cifuentes}, {Coimbra Araujo}, {Conforti}, {Conte}, {Contreras}, {Cortina},
  {Costa}, {Costantini}, {Cotter}, {Cristofari}, {Cuevas}, {Curtis-Ginsberg},
  {D'Amico}, {D'Ammando}, {Dalchenko}, {Dazzi}, {de Bony de Lavergne}, {De
  Caprio}, {De Frondat Laadim}, {de Gouveia Dal Pino}, {De Lotto}, {De Lucia},
  {De Martino}, {de Menezes}, {de Naurois}, {De Simone}, {de Souza}, {del
  Valle}, {Delagnes}, {Delgado Giler}, {Delgado}, {Dell'aiera}, {della Volpe},
  {Depaoli}, {Di Girolamo}, {Di Piano}, {Di Pierro}, {Di Tria}, {Di Venere},
  {Diebold}, {Djannati-Ata{\"\i}}, {Djuvsland}, {Dominik}, {Donini}, {Dorner},
  {D{\"o}rner}, {Doro}, {dos Anjos}, {Dournaux}, {Duangchan}, {Dubos},
  {Dumora}, {Dwarkadas}, {Ebr}, {Eckner}, {Egberts}, {Einecke}, {Els{\"a}sser},
  {Emery}, {Escobar Godoy}, {Escudero}, {Esposito}, {Ettori}, {Evoli},
  {Falceta-Goncalves}, {Fallah Ramazani}, {Fattorini}, {Faure}, {Fedorova},
  {Fegan}, {Feijen}, {Feng}, {Ferrand}, {Ferrarotto}, {Fiandrini}, {Fiasson},
  {Filipovic}, {Fioretti}, {Foffano}, {Font Guiteras}, {Fontaine}, {Fr{\"o}se},
  {Fukazawa}, {Fukui}, {Gaggero}, {Galanti}, {Gallozzi}, {Gammaldi},
  {Garczarczyk}, {Gasbarra}, {Gasparrini}, {Gaug}, {Ghalumyan}, {Gianotti},
  {Giarrusso}, {Giesbrecht}, {Giglietto}, {Giordano}, {Glicenstein},
  {G{\"o}ksu}, {Goldoni}, {Gonz{\'a}lez}, {Gonz{\'a}lez}, {Goulart Coelho},
  {Granot}, {Grau}, {Gr{\'e}aux}, {Green}, {Green}, {Grenier}, {Grolleron},
  {Grube}, {Gueta}, {Hackfeld}, {Hadasch}, {Hamal}, {Hanlon}, {Hara}, {Harvey},
  {Hassan}, {Heckmann}, {Heller}, {Hern{\'a}ndez Cadena}, {Hervet}, {Hie},
  {Hiroshima}, {Hnatyk}, {Hnatyk}, {Hoang}, {Hoffmann}, {Hofmann}, {Holder},
  {Horan}, {Horvath}, {Hrupec122}, {H{\"u}tten}, {Iarlori}, {Inada},
  {Incardona}, {Inoue}, {Iocco}, {Iori}, {Jamrozy}, {Janecek}, {Jankowsky},
  {Jarnot}, {Jean}, {Jim{\'e}nez Mart{\'\i}nez}, {Jin}, {Juramy-Gilles},
  {Jurysek}, {Kagaya}, {Kantzas}, {Karas}, {Katagiri}, {Kataoka}, {Kaufmann},
  {Kerszberg}, {Kh{\'e}lifi}, {Kissmann}, {Kleiner}, {Kluge}, {Klu{\'z}niak},
  {Kn{\"o}dlseder}, {Kobayashi}, {Kohri}, {Komin}, {Kornecki}, {Kosack},
  {Kowal}, {Kubo}, {Kushida}, {La Barbera}, {La Palombara}, {L{\'a}inez},
  {Lamastra}, {Lapington}, {Laporte}, {Lazarevi{\'c}}, {Leitgeb},
  {Lemoine-Goumard}, {Lenain}, {Leone}, {Leto}, {Leuschner}, {Lindfors},
  {Linhoff}, {Liodakis}, {Lombardi}, {Longo}, {L{\'o}pez-Coto},
  {L{\'o}pez-Moya}, {L{\'o}pez-Oramas}, {Loporchio}, {Luque-Escamilla},
  {Macias}, {Mackey}, {Majumdar}, {Malyshev}, {Mandat}, {Manganaro},
  {Manic{\`o}}, {Mariotti}, {Markoff}, {M{\'a}rquez}, {Marquez}, {Marsella},
  {Mart{\'\i}nez}, {Mart{\'\i}nez}, {Martinez}, {Marty}, {Mas-Aguilar},
  {Mastropietro}, {Maurin}, {Mazin}, {Melkumyan}, {Mello}, {Meunier}, {Meyer},
  {Meyer}, {Miceli}, {Michailidis}, {Micha{\l}owski}, {Miener}, {Miranda},
  {Mitchell}, {Mizote}, {Mizuno}, {Moderski}, {Molero}, {Molfese}, {Molina},
  {Montaruli}, {Morcuende}, {Morik}, {Morlino}, {Morselli}, {Moulin}, {Moya
  Zamanillo}, {Munari}, {Murach}, {Muraczewski}, {Muraishi}, {Nagataki},
  {Nakamori}, {Nemmen}, {Neyroud}, {Nickel}, {Niemiec}, {Nieto}, {Nievas
  Rosillo}, {Niko{\l}ajuk}, {Nishijima}, {Noda}, {Nosek}, {Novotny}, {Nozaki},
  {O'Brien}, {Ohishi}, {Ohtani}, {Okumura}, {Olive}, {Olmi}, {Ong}, {Orienti},
  {Orito}, {Orlandini}, {Orlando}, {Ostrowski}, {Oya}, {Pagliaro},
  {Palatiello}, {Panebianco}, {Paneque}, {Pantaleo}, {Paoletti}, {Paredes},
  {Parmiggiani}, {Patel}, {Patricelli}, {Pavlovi{\'c}}, {Pech}, {Pecimotika},
  {Pensec}, {Peresano}, {P{\'e}rez-Romero}, {Peron}, {Persic}, {Petrucci},
  {Petruk}, {Piano}, {Pierre}, {Pietropaolo}, {Pintore}, {Pirola}, {Pita},
  {Plard}, {Podobnik}, {Pohl}, {Polo}, {Pons}, {Ponti}, {Prandini}, {Prast},
  {Principe}, {Priyadarshi}, {Produit}, {Pueschel}, {P{\"u}hlhofer}, {Pumo},
  {Punch}, {Queiroz}, {Quirrenbach}, {Rain{\`o}}, {Rando}, {Razzaque},
  {Recchia}, {Regeard}, {Reichherzer}, {Reimer}, {Reimer}, {Reisenegger},
  {Rhode}, {Ribeiro}, {Rib{\'o}}, {Richtler}, {Rico}, {Rieger}, {Righi},
  {Riitano}, {Rizi}, {Roache}, {Rodriguez Fernandez},
  {Rodr{\'\i}guez-V{\'a}zquez}, {Romano}, {Romeo}, {Rosado}, {Rosales de Leon},
  {Rowell}, {Rudak}, {Rulten}, {Russo}, {Sadeh}, {Saha}, {Saito}, {Salzmann},
  {Sanchez}, {S{\'a}nchez-Conde}, {Sangiorgi}, {Sano}, {Santander},
  {Santangelo}, {Santos-Lima}, {Sanuy}, {{\v{S}}ari{\'c}}, {Sarkar}, {Sarkar},
  {Satalecka}, {Saturni}, {Savchenko}, {Scherer}, {Schipani}, {Schleicher},
  {Schubert}, {Schussler}, {Schwanke}, {Schwefer}, {Seglar Arroyo}, {Seiji},
  {Semikoz}, {Sergijenko}, {Servillat}, {Sguera}, {Shang}, {Sharma},
  {Siejkowski}, {Sinha}, {Siqueira}, {Sliusar}, {Slowikowska}, {Sol},
  {Specovius}, {Spencer}, {Spiga}, {Stamerra}, {Stani{\v{c}}}, {Starecki},
  {Starling}, {Stawarz}, {Steppa}, {Stolarczyk}, {Stri{\v{s}}kovi{\'c}},
  {Suda}, {Suomij{\"a}rvi}, {Tajima}, {Tak}, {Takahashi}, {Takeishi}, {Tanaka},
  {Tavernier}, {Tejedor}, {Terauchi}, {Terrier}, {Teshima}, {Tian}, {Tibaldo},
  {Tibolla}, {Torradeflot}, {Torres}, {Torresi}, {Tosti}, {Tosti}, {Tothill},
  {Toussenel}, {Touzard}, {Tramacere}, {Travnicek}, {Tripodo}, {Truzzi},
  {Tsiahina}, {Tutone}, {Vacula}, {Vallage}, {Vallania}, {van Eldik}, {van
  Scherpenberg}, {Vandenbroucke}, {Vassiliev}, {V{\'a}zquez Acosta}, {Vecchi},
  {Ventura}, {Vercellone}, {Verna}, {Viana}, {Viaux}, {Vigliano}, {Vigorito},
  {Vitale}, {Vodeb}, {Voisin}, {Vorobiov}, {Voutsinas}, {Vovk}, {Vuillaume},
  {Wagner}, {Walter}, {Wechakama}, {White}, {Wierzcholska}, {Will}, {Williams},
  {Wohlleben}, {Wolter}, {Yamamoto}, {Yamazaki}, {Yoshida}, {Yoshikoshi},
  {Zacharias}, {Zaharijas}, {Zavrtanik}, {Zavrtanik}, {Zdziarski}, {Zech},
  {Zhdanov}, {{\v{Z}}ivec}, {Zuriaga-Puig}, \& {De la Torre
  Luque}}]{2023arXiv230903712C}
{Cherenkov Telescope Array Consortium}, T., {:}, {Abe}, K., {et~al.} 2023,
  arXiv e-prints, arXiv:2309.03712

\bibitem[{{Chevalier} \& {Liang}(1989)}]{chevalier_apj_344_1989}
{Chevalier}, R.~A. \& {Liang}, E.~P. 1989, \apj, 344, 332

\bibitem[{{Chevalier} \& {Luo}(1994)}]{chevalier_apj_421_1994}
{Chevalier}, R.~A. \& {Luo}, D. 1994, \apj, 421, 225

\bibitem[{{Chiotellis} {et~al.}(2020){Chiotellis}, {Boumis}, \&
  {Spetsieri}}]{chiotellis_galax_8_2020}
{Chiotellis}, A., {Boumis}, P., \& {Spetsieri}, Z.~T. 2020, Galaxies, 8, 38

\bibitem[{{Chiotellis} {et~al.}(2021){Chiotellis}, {Boumis}, \&
  {Spetsieri}}]{chiotellis_mnras_502_2021}
{Chiotellis}, A., {Boumis}, P., \& {Spetsieri}, Z.~T. 2021, \mnras, 502, 176

\bibitem[{{Chiotellis} {et~al.}(2013){Chiotellis}, {Kosenko}, {Schure}, {Vink},
  \& {Kaastra}}]{chiotellis_mnras_435_2013}
{Chiotellis}, A., {Kosenko}, D., {Schure}, K.~M., {Vink}, J., \& {Kaastra},
  J.~S. 2013, \mnras, 435, 1659

\bibitem[{{Chiotellis} {et~al.}(2012){Chiotellis}, {Schure}, \&
  {Vink}}]{chiotellis_aa_537_2012}
{Chiotellis}, A., {Schure}, K.~M., \& {Vink}, J. 2012, \aap, 537, A139

\bibitem[{{de la Chevroti{\`e}re} {et~al.}(2014){de la Chevroti{\`e}re},
  {St-Louis}, {Moffat}, \& {MiMeS Collaboration}}]{chevrotiere_apj_781_2014}
{de la Chevroti{\`e}re}, A., {St-Louis}, N., {Moffat}, A.~F.~J., \& {MiMeS
  Collaboration}. 2014, \apj, 781, 73

\bibitem[{{Decin} {et~al.}(2012){Decin}, { }, {Royer}, {Van Marle},
  {Vandenbussche}, {Ladjal}, {Kerschbaum}, {Ottensamer}, {Barlow}, {Blommaert},
  {Gomez}, {Groenewegen}, {Lim}, {Swinyard}, {Waelkens}, \&
  {Tielens}}]{2ssssssssssssssss}
{Decin}, L., { }, N.~L.~J., {Royer}, P., {et~al.} 2012, \aap, 548, A113

\bibitem[{{Eggenberger} {et~al.}(2008){Eggenberger}, {Meynet}, {Maeder},
  {Hirschi}, {Charbonnel}, {Talon}, \& {Ekstr{\"o}m}}]{2008Ap&SS.316...43E}
{Eggenberger}, P., {Meynet}, G., {Maeder}, A., {et~al.} 2008, \apss, 316, 43

\bibitem[{{Ekstr{\"o}m} {et~al.}(2012){Ekstr{\"o}m}, {Georgy}, {Eggenberger},
  {Meynet}, {Mowlavi}, {Wyttenbach}, {Granada}, {Decressin}, {Hirschi},
  {Frischknecht}, {Charbonnel}, \& {Maeder}}]{ekstroem_aa_537_2012}
{Ekstr{\"o}m}, S., {Georgy}, C., {Eggenberger}, P., {et~al.} 2012, \aap, 537,
  A146

\bibitem[{{El Mellah} {et~al.}(2020){El Mellah}, {Bolte}, {Decin}, {Homan}, \&
  {Keppens}}]{elmellah_aa_637_2020}
{El Mellah}, I., {Bolte}, J., {Decin}, L., {Homan}, W., \& {Keppens}, R. 2020,
  \aap, 637, A91

\bibitem[{{Eldridge} {et~al.}(2006){Eldridge}, {Genet}, {Daigne}, \&
  {Mochkovitch}}]{eldridge_mnras_367_2006}
{Eldridge}, J.~J., {Genet}, F., {Daigne}, F., \& {Mochkovitch}, R. 2006,
  \mnras, 367, 186

\bibitem[{{Fang} {et~al.}(2017){Fang}, {Yu}, \& {Zhang}}]{fang_mnras_464_2017}
{Fang}, J., {Yu}, H., \& {Zhang}, L. 2017, \mnras, 464, 940

\bibitem[{{Ferrand} \& {Safi-Harb}(2012)}]{ferrand_adspr_49_2012}
{Ferrand}, G. \& {Safi-Harb}, S. 2012, Advances in Space Research, 49, 1313

\bibitem[{{Fossati} {et~al.}(2015){Fossati}, {Castro}, {Morel}, {Langer},
  {Briquet}, {Carroll}, {Hubrig}, {Nieva}, {Oskinova}, {Przybilla},
  {Schneider}, {Sch{\"o}ller}, {Sim{\'o}n-D{\'\i}az}, {Ilyin}, {de Koter},
  {Reisenegger}, \& {Sana}}]{fossati_aa_574_2015}
{Fossati}, L., {Castro}, N., {Morel}, T., {et~al.} 2015, \aap, 574, A20

\bibitem[{{Friend} \& {Abbott}(1986)}]{1986ApJ...311..701F}
{Friend}, D.~B. \& {Abbott}, D.~C. 1986, \apj, 311, 701

\bibitem[{{Gabler} {et~al.}(2021){Gabler}, {Wongwathanarat}, \&
  {Janka}}]{gabler_mnras_502_2021}
{Gabler}, M., {Wongwathanarat}, A., \& {Janka}, H.-T. 2021, \mnras, 502, 3264

\bibitem[{{Gaensler}(1999)}]{gaensler_phd_1999}
{Gaensler}, B.~M. 1999, PhD thesis, University of Sydney

\bibitem[{{Gaensler} {et~al.}(1998){Gaensler}, {Green}, \&
  {Manchester}}]{gaensler_mnras_299_1998}
{Gaensler}, B.~M., {Green}, A.~J., \& {Manchester}, R.~N. 1998, \mnras, 299,
  812

\bibitem[{{Georgy} {et~al.}(2011){Georgy}, {Meynet}, \&
  {Maeder}}]{Georgy_a__527_2011}
{Georgy}, C., {Meynet}, G., \& {Maeder}, A. 2011, \aap, 527, A52

\bibitem[{{Ghisellini}(2013)}]{ghiselini_book}
{Ghisellini}, G. 2013, { Radiative Processes in High Energy Astrophysics }
  (Springer Cham)

\bibitem[{{Gilkis} {et~al.}(2016){Gilkis}, {Soker}, \&
  {Papish}}]{2016ApJ...826..178G}
{Gilkis}, A., {Soker}, N., \& {Papish}, O. 2016, \apj, 826, 178

\bibitem[{{Gull} \& {Sofia}(1979)}]{gull_apj_230_1979}
{Gull}, T.~R. \& {Sofia}, S. 1979, \apj, 230, 782

\bibitem[{{Gvaramadze}(1999)}]{gvaramadez_1999}
{Gvaramadze}, V.~V. 1999, Odessa Astronomical Publications, 12, 117

\bibitem[{{Gvaramadze} {et~al.}(2015){Gvaramadze}, {Kniazev}, {Bestenlehner},
  {Bodensteiner}, {Langer}, {Greiner}, {Grebel}, {Berdnikov}, \&
  {Beletsky}}]{Gvaramadze_mnras_454_2015}
{Gvaramadze}, V.~V., {Kniazev}, A.~Y., {Bestenlehner}, J.~M., {et~al.} 2015,
  \mnras, 454, 219

\bibitem[{{Hamann} {et~al.}(2019){Hamann}, {Gr{\"a}fener}, {Liermann},
  {Hainich}, {Sander}, {Shenar}, {Ramachandran}, {Todt}, \&
  {Oskinova}}]{hamman_aa_625_2019}
{Hamann}, W.~R., {Gr{\"a}fener}, G., {Liermann}, A., {et~al.} 2019, \aap, 625,
  A57

\bibitem[{Harten {et~al.}(1983)Harten, Lax, \& van Leer}]{hll_ref}
Harten, A., Lax, P.~D., \& van Leer, B. 1983, SIAM Review, 25, 35

\bibitem[{{Heger} {et~al.}(2000){Heger}, {Langer}, \&
  {Woosley}}]{heger_apj_apj_2000}
{Heger}, A., {Langer}, N., \& {Woosley}, S.~E. 2000, \apj, 528, 368

\bibitem[{{Henney} {et~al.}(2009){Henney}, {Arthur}, {de Colle}, \&
  {Mellema}}]{henney_mnras_398_2009}
{Henney}, W.~J., {Arthur}, S.~J., {de Colle}, F., \& {Mellema}, G. 2009,
  \mnras, 398, 157

\bibitem[{{Herbst} {et~al.}(2020){Herbst}, {Scherer}, {Ferreira}, {Baalmann},
  {Engelbrecht}, {Fichtner}, {Kleimann}, {Strauss}, {Moeketsi}, \&
  {Mohamed}}]{herbst_apj_897_2020}
{Herbst}, K., {Scherer}, K., {Ferreira}, S. E.~S., {et~al.} 2020, \apjl, 897,
  L27

\bibitem[{{Hubrig} {et~al.}(2016){Hubrig}, {Scholz}, {Hamann}, {Sch{\"o}ller},
  {Ignace}, {Ilyin}, {Gayley}, \& {Oskinova}}]{hubrig_mnras_458_2016}
{Hubrig}, S., {Scholz}, K., {Hamann}, W.~R., {et~al.} 2016, \mnras, 458, 3381

\bibitem[{{Hummer}(1994)}]{hummer_mnras_268_1994}
{Hummer}, D.~G. 1994, \mnras, 268, 109

\bibitem[{{Kaplan} \& {Soker}(2020)}]{2020MNRAS.492.3013K}
{Kaplan}, N. \& {Soker}, N. 2020, \mnras, 492, 3013

\bibitem[{{Katsuda} {et~al.}(2018){Katsuda}, {Takiwaki}, {Tominaga}, {Moriya},
  \& {Nakamura}}]{katsuda_apj_863_2018}
{Katsuda}, S., {Takiwaki}, T., {Tominaga}, N., {Moriya}, T.~J., \& {Nakamura},
  K. 2018, \apj, 863, 127

\bibitem[{{Kervella} {et~al.}(2018){Kervella}, {Decin}, {Richards}, {Harper},
  {McDonald}, {O'Gorman}, {Montarg{\`e}s}, {Homan}, \&
  {Ohnaka}}]{kervella_aa_609_2018}
{Kervella}, P., {Decin}, L., {Richards}, A. M.~S., {et~al.} 2018, \aap, 609,
  A67

\bibitem[{{Kesteven} \& {Caswell}(1987)}]{kesteven_aa_183_1987}
{Kesteven}, M.~J. \& {Caswell}, J.~L. 1987, \aap, 183, 118

\bibitem[{{Kroupa}(2001)}]{kroupa_mnras_322_2001}
{Kroupa}, P. 2001, \mnras, 322, 231

\bibitem[{{Langer}(2012)}]{langer_araa_50_2012}
{Langer}, N. 2012, \araa, 50, 107

\bibitem[{{Madura} {et~al.}(2013){Madura}, {Gull}, {Okazaki}, {Russell},
  {Owocki}, {Groh}, {Corcoran}, {Hamaguchi}, \&
  {Teodoro}}]{Madura_mnras_436_2013}
{Madura}, T.~I., {Gull}, T.~R., {Okazaki}, A.~T., {et~al.} 2013, \mnras, 436,
  3820

\bibitem[{{Martayan} {et~al.}(2016){Martayan}, {Lobel}, {Baade}, {Mehner},
  {Rivinius}, {Boffin}, {Girard}, {Mawet}, {Montagnier}, {Blomme}, {Kervella},
  {Sana}, {{\v{S}}tefl}, {Zorec}, {Lacour}, {Le Bouquin}, {Martins},
  {M{\'e}rand}, {Patru}, {Selman}, \& {Fr{\'e}mat}}]{martayan_aa_587_2016}
{Martayan}, C., {Lobel}, A., {Baade}, D., {et~al.} 2016, \aap, 587, A115

\bibitem[{{Meyer}(2021)}]{meyer_mnras_507_2021}
{Meyer}, D.~M.~A. 2021, \mnras, 507, 4697

\bibitem[{{Meyer} {et~al.}(2015){Meyer}, {Langer}, {Mackey}, {Vel{\'a}zquez},
  \& {Gusdorf}}]{meyer_mnras_450_2015}
{Meyer}, D.~M.-A., {Langer}, N., {Mackey}, J., {Vel{\'a}zquez}, P.~F., \&
  {Gusdorf}, A. 2015, \mnras, 450, 3080

\bibitem[{{Meyer} {et~al.}(2014){Meyer}, {Mackey}, {Langer}, {Gvaramadze},
  {Mignone}, {Izzard}, \& {Kaper}}]{meyer_2014bb}
{Meyer}, D.~M.-A., {Mackey}, J., {Langer}, N., {et~al.} 2014, \mnras, 444, 2754

\bibitem[{{Meyer} {et~al.}(2024){Meyer}, {Meliani}, {Vel{\'a}zquez}, {Pohl}, \&
  {Torres}}]{meyer_mnras_527_2024}
{Meyer}, D.~M.~A., {Meliani}, Z., {Vel{\'a}zquez}, P.~F., {Pohl}, M., \&
  {Torres}, D.~F. 2024, \mnras, 527, 5514

\bibitem[{{Meyer} {et~al.}(2021{\natexlab{a}}){Meyer}, {Mignone}, {Petrov},
  {Scherer}, {Vel{\'a}zquez}, \& {Boumis}}]{meyer_mnras_506_2021}
{Meyer}, D.~M.~A., {Mignone}, A., {Petrov}, M., {et~al.} 2021{\natexlab{a}},
  \mnras, 506, 5170

\bibitem[{{Meyer} {et~al.}(2020){Meyer}, {Petrov}, \&
  {Pohl}}]{meyer_mnras_493_2020}
{Meyer}, D.~M.~A., {Petrov}, M., \& {Pohl}, M. 2020, \mnras, 493, 3548

\bibitem[{{Meyer} {et~al.}(2023){Meyer}, {Pohl}, {Petrov}, \&
  {Egberts}}]{meyer_mnras_521_2023}
{Meyer}, D.~M.~A., {Pohl}, M., {Petrov}, M., \& {Egberts}, K. 2023, \mnras,
  521, 5354

\bibitem[{{Meyer} {et~al.}(2021{\natexlab{b}}){Meyer}, {Pohl}, {Petrov}, \&
  {Oskinova}}]{meyer_mnras_502_2021}
{Meyer}, D.~M.~A., {Pohl}, M., {Petrov}, M., \& {Oskinova}, L.
  2021{\natexlab{b}}, \mnras, 502, 5340

\bibitem[{{Meyer} {et~al.}(2022){Meyer}, {Vel{\'a}zquez}, {Petruk},
  {Chiotellis}, {Pohl}, {Camps-Fari{\~n}a}, {Petrov}, {Reynoso}, {Toledo-Roy},
  {Schneiter}, {Castellanos-Ram{\'\i}rez}, \&
  {Esquivel}}]{mnras_mnras_515_2022}
{Meyer}, D.~M.~A., {Vel{\'a}zquez}, P.~F., {Petruk}, O., {et~al.} 2022, \mnras,
  515, 594

\bibitem[{{Meyer} {et~al.}(2017){Meyer}, {Vorobyov}, {Kuiper}, \&
  {Kley}}]{meyer_mnras_464_2017}
{Meyer}, D.~M.-A., {Vorobyov}, E.~I., {Kuiper}, R., \& {Kley}, W. 2017, \mnras,
  464, L90

\bibitem[{Mignone(2014)}]{MIGNONE2014784}
Mignone, A. 2014, Journal of Computational Physics, 270, 784

\bibitem[{{Mignone} {et~al.}(2007){Mignone}, {Bodo}, {Massaglia}, {Matsakos},
  {Tesileanu}, {Zanni}, \& {Ferrari}}]{mignone_apj_170_2007}
{Mignone}, A., {Bodo}, G., {Massaglia}, S., {et~al.} 2007, \apjs, 170, 228

\bibitem[{{Mignone} {et~al.}(2012){Mignone}, {Zanni}, {Tzeferacos}, {van
  Straalen}, {Colella}, \& {Bodo}}]{migmone_apjs_198_2012}
{Mignone}, A., {Zanni}, C., {Tzeferacos}, P., {et~al.} 2012, \apjs, 198, 7

\bibitem[{{Mohamed} {et~al.}(2012){Mohamed}, {Mackey}, \&
  {Langer}}]{mohamed_aa_541_2012}
{Mohamed}, S., {Mackey}, J., \& {Langer}, N. 2012, \aap, 541, A1

\bibitem[{{M{\"u}ller} {et~al.}(2012){M{\"u}ller}, {Janka}, \&
  {Wongwathanarat}}]{2012A&A...537A..63M}
{M{\"u}ller}, E., {Janka}, H.~T., \& {Wongwathanarat}, A. 2012, \aap, 537, A63

\bibitem[{{Noriega-Crespo} {et~al.}(1997){Noriega-Crespo}, {van Buren}, {Cao},
  \& {Dgani}}]{noriegacrespo_aj_114_1997}
{Noriega-Crespo}, A., {van Buren}, D., {Cao}, Y., \& {Dgani}, R. 1997, \aj,
  114, 837

\bibitem[{{Orlando} {et~al.}(2012){Orlando}, {Bocchino}, {Miceli}, {Petruk}, \&
  {Pumo}}]{orlando_apj_749_2012}
{Orlando}, S., {Bocchino}, F., {Miceli}, M., {Petruk}, O., \& {Pumo}, M.~L.
  2012, \apj, 749, 156

\bibitem[{{Orlando} {et~al.}(2007){Orlando}, {Bocchino}, {Reale}, {Peres}, \&
  {Petruk}}]{orlando_aa_470_2007}
{Orlando}, S., {Bocchino}, F., {Reale}, F., {Peres}, G., \& {Petruk}, O. 2007,
  \aap, 470, 927

\bibitem[{{Orlando} {et~al.}(2019){Orlando}, {Miceli}, {Petruk}, {Ono},
  {Nagataki}, {Aloy}, {Mimica}, {Lee}, {Bocchino}, {Peres}, \&
  {Guarrasi}}]{orlando_aa_622_2019}
{Orlando}, S., {Miceli}, M., {Petruk}, O., {et~al.} 2019, \aap, 622, A73

\bibitem[{{Orlando} {et~al.}(2020){Orlando}, {Ono}, {Nagataki}, {Miceli},
  {Umeda}, {Ferrand}, {Bocchino}, {Petruk}, {Peres}, {Takahashi}, \&
  {Yoshida}}]{orlando_aa_636_2020}
{Orlando}, S., {Ono}, M., {Nagataki}, S., {et~al.} 2020, \aap, 636, A22

\bibitem[{{Orlando} {et~al.}(2021){Orlando}, {Wongwathanarat}, {Janka},
  {Miceli}, {Ono}, {Nagataki}, {Bocchino}, \& {Peres}}]{orlando_aa_645_2021}
{Orlando}, S., {Wongwathanarat}, A., {Janka}, H.~T., {et~al.} 2021, \aap, 645,
  A66

\bibitem[{{Osterbrock}(1989)}]{osterbrock_1989}
{Osterbrock}, D.~E. 1989, University Science Books, Mill Valley, CA

\bibitem[{{Papish} \& {Soker}(2011)}]{2011MNRAS.416.1697P}
{Papish}, O. \& {Soker}, N. 2011, \mnras, 416, 1697

\bibitem[{{Papish} \& {Soker}(2014)}]{2014MNRAS.438.1027P}
{Papish}, O. \& {Soker}, N. 2014, \mnras, 438, 1027

\bibitem[{{Parker}(1958)}]{parker_paj_128_1958}
{Parker}, E.~N. 1958, \apj, 128, 664

\bibitem[{{Parkin} {et~al.}(2011){Parkin}, {Pittard}, {Corcoran}, \&
  {Hamaguchi}}]{parkin_apj_726_2011}
{Parkin}, E.~R., {Pittard}, J.~M., {Corcoran}, M.~F., \& {Hamaguchi}, K. 2011,
  \apj, 726, 105

\bibitem[{{Peri} {et~al.}(2012){Peri}, {Benaglia}, {Brookes}, {Stevens}, \&
  {Isequilla}}]{peri_aa_538_2012}
{Peri}, C.~S., {Benaglia}, P., {Brookes}, D.~P., {Stevens}, I.~R., \&
  {Isequilla}, N.~L. 2012, \aap, 538, A108

\bibitem[{{Peri} {et~al.}(2015){Peri}, {Benaglia}, \&
  {Isequilla}}]{peri_aa_578_2015}
{Peri}, C.~S., {Benaglia}, P., \& {Isequilla}, N.~L. 2015, \aap, 578, A45

\bibitem[{{Petruk} {et~al.}(2009){Petruk}, {Dubner}, {Castelletti}, {Bocchino},
  {Iakubovskyi}, {Kirsch}, {Miceli}, {Orlando}, \&
  {Telezhinsky}}]{petruk_393_mnras_2009}
{Petruk}, O., {Dubner}, G., {Castelletti}, G., {et~al.} 2009, \mnras, 393, 1034

\bibitem[{{Pogorelov} \& {Matsuda}(2000)}]{pogolerov_aa_354_2000}
{Pogorelov}, N.~V. \& {Matsuda}, T. 2000, \aap, 354, 697

\bibitem[{{Pogorelov} \& {Semenov}(1997)}]{pogolerov_aa_321_1997}
{Pogorelov}, N.~V. \& {Semenov}, A.~Y. 1997, \aap, 321, 330

\bibitem[{Powell(1997)}]{Powell1997}
Powell, K.~G. 1997, An Approximate Riemann Solver for Magnetohydrodynamics, ed.
  M.~Y. Hussaini, B.~van Leer, \& J.~Van~Rosendale (Berlin, Heidelberg:
  Springer Berlin Heidelberg), 570--583

\bibitem[{{Przybilla} {et~al.}(2016){Przybilla}, {Fossati}, {Hubrig}, {Nieva},
  {J{\"a}rvinen}, {Castro}, {Sch{\"o}ller}, {Ilyin}, {Butler}, {Schneider},
  {Oskinova}, {Morel}, {Langer}, {de Koter}, \& {BOB
  Collaboration}}]{przybilla_aa_587_2016}
{Przybilla}, N., {Fossati}, L., {Hubrig}, S., {et~al.} 2016, \aap, 587, A7

\bibitem[{{Rozyczka} \& {Franco}(1996)}]{rozyczka_apj_469_1996}
{Rozyczka}, M. \& {Franco}, J. 1996, \apjl, 469, L127

\bibitem[{{Sana} {et~al.}(2012){Sana}, {de Mink}, {de Koter}, {Langer},
  {Evans}, {Gieles}, {Gosset}, {Izzard}, {Le Bouquin}, \&
  {Schneider}}]{sana_sci_337_2012}
{Sana}, H., {de Mink}, S.~E., {de Koter}, A., {et~al.} 2012, Science, 337, 444

\bibitem[{{Scherer} {et~al.}(2020){Scherer}, {Baalmann}, {Fichtner},
  {Kleimann}, {Bomans}, {Weis}, {Ferreira}, \&
  {Herbst}}]{scherer_mnras_493_2020}
{Scherer}, K., {Baalmann}, L.~R., {Fichtner}, H., {et~al.} 2020, \mnras, 493,
  4172

\bibitem[{{Shimizu} {et~al.}(2001){Shimizu}, {Ebisuzaki}, {Sato}, \&
  {Yamada}}]{shimizu_apj_552_2001}
{Shimizu}, T.~M., {Ebisuzaki}, T., {Sato}, K., \& {Yamada}, S. 2001, \apj, 552,
  756

\bibitem[{{Shishkin} \& {Soker}(2023)}]{shishkin_mnras_522_2023}
{Shishkin}, D. \& {Soker}, N. 2023, \mnras, 522, 438

\bibitem[{{Smartt}(2009)}]{smartt_araa_47_2009}
{Smartt}, S.~J. 2009, \araa, 47, 63

\bibitem[{{Soker}(2021)}]{soker_apj_906_2021}
{Soker}, N. 2021, \apj, 906, 1

\bibitem[{{Soker}(2022{\natexlab{a}})}]{2022RAA....22i5007S}
{Soker}, N. 2022{\natexlab{a}}, Research in Astronomy and Astrophysics, 22,
  095007

\bibitem[{{Soker}(2022{\natexlab{b}})}]{2022MNRAS.516.4942S}
{Soker}, N. 2022{\natexlab{b}}, \mnras, 516, 4942

\bibitem[{{Soker}(2022{\natexlab{c}})}]{2022RAA....22l2003S}
{Soker}, N. 2022{\natexlab{c}}, Research in Astronomy and Astrophysics, 22,
  122003

\bibitem[{{Soker}(2023{\natexlab{a}})}]{soker_raa_2023}
{Soker}, N. 2023{\natexlab{a}}, Research in Astronomy and Astrophysics, 23,
  115017

\bibitem[{{Soker}(2023{\natexlab{b}})}]{soker_raa_2023_23l1001S}
{Soker}, N. 2023{\natexlab{b}}, Research in Astronomy and Astrophysics, 23,
  121001

\bibitem[{{Soker} \& {Kaplan}(2021)}]{soker_apj_907_2021}
{Soker}, N. \& {Kaplan}, N. 2021, \apj, 907, 120

\bibitem[{{Sukhbold} {et~al.}(2016){Sukhbold}, {Ertl}, {Woosley}, {Brown}, \&
  {Janka}}]{sukhbold_apj_821_2016}
{Sukhbold}, T., {Ertl}, T., {Woosley}, S.~E., {Brown}, J.~M., \& {Janka}, H.~T.
  2016, \apj, 821, 38

\bibitem[{{Temim} {et~al.}(2015){Temim}, {Slane}, {Kolb}, {Blondin}, {Hughes},
  \& {Bucciantini}}]{temim_apj_808_2015}
{Temim}, T., {Slane}, P., {Kolb}, C., {et~al.} 2015, \apj, 808, 100

\bibitem[{{Temim} {et~al.}(2017){Temim}, {Slane}, {Plucinsky}, {Gelfand},
  {Castro}, \& {Kolb}}]{temim_apj_851_2017}
{Temim}, T., {Slane}, P., {Plucinsky}, P.~P., {et~al.} 2017, \apj, 851, 128

\bibitem[{{Temim} {et~al.}(2022){Temim}, {Slane}, {Raymond}, {Patnaude},
  {Murray}, {Ghavamian}, {Renzo}, \& {Jacovich}}]{temim_apj_932_2022}
{Temim}, T., {Slane}, P., {Raymond}, J.~C., {et~al.} 2022, \apj, 932, 26

\bibitem[{{Truelove} \& {McKee}(1999)}]{truelove_apjs_120_1999}
{Truelove}, J.~K. \& {McKee}, C.~F. 1999, \apjs, 120, 299

\bibitem[{{Uchida} {et~al.}(2009){Uchida}, {Tsunemi}, {Katsuda}, {Kimura}, \&
  {Kosugi}}]{uchida_pasj_61_2009}
{Uchida}, H., {Tsunemi}, H., {Katsuda}, S., {Kimura}, M., \& {Kosugi}, H. 2009,
  \pasj, 61, 301

\bibitem[{{Uyan{\i}ker} {et~al.}(2004){Uyan{\i}ker}, {Reich}, {Yar}, \&
  {F{\"u}rst}}]{uyaniker_426_aa_2004}
{Uyan{\i}ker}, B., {Reich}, W., {Yar}, A., \& {F{\"u}rst}, E. 2004, \aap, 426,
  909

\bibitem[{{Vaidya} {et~al.}(2018){Vaidya}, {Mignone}, {Bodo}, {Rossi}, \&
  {Massaglia}}]{vaidya_apj_865_2018}
{Vaidya}, B., {Mignone}, A., {Bodo}, G., {Rossi}, P., \& {Massaglia}, S. 2018,
  \apj, 865, 144

\bibitem[{{van Buren} \& {McCray}(1988)}]{buren_apj_329_1988}
{van Buren}, D. \& {McCray}, R. 1988, \apjl, 329, L93

\bibitem[{{van Marle} {et~al.}(2014){van Marle}, {Decin}, \&
  {Meliani}}]{vanmarle_aa_561_2014}
{van Marle}, A.~J., {Decin}, L., \& {Meliani}, Z. 2014, \aap, 561, A152

\bibitem[{{van Marle} {et~al.}(2015){van Marle}, {Meliani}, \&
  {Marcowith}}]{vanmarle_584_aa_2015}
{van Marle}, A.~J., {Meliani}, Z., \& {Marcowith}, A. 2015, \aap, 584, A49

\bibitem[{{van Marle} {et~al.}(2010){van Marle}, {Smith}, {Owocki}, \& {van
  Veelen}}]{vanmarle_mnras_407_2010}
{van Marle}, A.~J., {Smith}, N., {Owocki}, S.~P., \& {van Veelen}, B. 2010,
  \mnras, 407, 2305

\bibitem[{{van Veelen} {et~al.}(2009){van Veelen}, {Langer}, {Vink},
  {Garc{\'{\i}}a-Segura}, \& {van Marle}}]{vanveelen_aa_50_2009}
{van Veelen}, B., {Langer}, N., {Vink}, J., {Garc{\'{\i}}a-Segura}, G., \& {van
  Marle}, A.~J. 2009, \aap, 503, 495

\bibitem[{{Vel{\'a}zquez} {et~al.}(2023){Vel{\'a}zquez}, {Meyer}, {Chiotellis},
  {Cruz-{\'A}lvarez}, {Schneiter}, {Toledo-Roy}, {Reynoso}, \&
  {Esquivel}}]{velazquez_mnras_519_2023}
{Vel{\'a}zquez}, P.~F., {Meyer}, D.~M.~A., {Chiotellis}, A., {et~al.} 2023,
  \mnras, 519, 5358

\bibitem[{{Vigh} {et~al.}(2011){Vigh}, {Vel{\'a}zquez}, {G{\'o}mez}, {Reynoso},
  {Esquivel}, \& {Matias Schneiter}}]{vigh_apj_727_2011}
{Vigh}, C.~D., {Vel{\'a}zquez}, P.~F., {G{\'o}mez}, D.~O., {et~al.} 2011, \apj,
  727, 32

\bibitem[{{Villagran} {et~al.}(2024){Villagran}, {G{\'o}mez}, {Vel{\'a}zquez},
  {Meyer}, {Chiotellis}, {Raga}, {Esquivel}, {Toledo-Roy}, {Vargas-Rojas}, \&
  {Schneiter}}]{2023arXiv230916410V}
{Villagran}, M.~A., {G{\'o}mez}, D.~O., {Vel{\'a}zquez}, P.~F., {et~al.} 2024,
  \mnras, 527, 1601

\bibitem[{{Vink}(2012)}]{vink_aarv_20_2012}
{Vink}, J. 2012, \aapr, 20, 49

\bibitem[{{Vink} {et~al.}(1996){Vink}, {Kaastra}, \&
  {Bleeker}}]{vink_aa_307_1996}
{Vink}, J., {Kaastra}, J.~S., \& {Bleeker}, J.~A.~M. 1996, \aap, 307, L41

\bibitem[{{Vink} {et~al.}(1997){Vink}, {Kaastra}, \&
  {Bleeker}}]{vink_aa_328_1997}
{Vink}, J., {Kaastra}, J.~S., \& {Bleeker}, J.~A.~M. 1997, \aap, 328, 628

\bibitem[{{Vishniac}(1994)}]{vishniac_apj_428_1994}
{Vishniac}, E.~T. 1994, \apj, 428, 186

\bibitem[{{Vlemmings} {et~al.}(2002){Vlemmings}, {Diamond}, \& {van
  Langevelde}}]{vlemmings_aa_394_2002}
{Vlemmings}, W.~H.~T., {Diamond}, P.~J., \& {van Langevelde}, H.~J. 2002, \aap,
  394, 589

\bibitem[{{Vlemmings} {et~al.}(2005){Vlemmings}, {van Langevelde}, \&
  {Diamond}}]{vlemmings_aa_434_2005}
{Vlemmings}, W.~H.~T., {van Langevelde}, H.~J., \& {Diamond}, P.~J. 2005, \aap,
  434, 1029

\bibitem[{{Wang} \& {Mazzali}(1992)}]{wang_nature_1992}
{Wang}, L. \& {Mazzali}, P.~A. 1992, \nat, 355, 58

\bibitem[{{Weaver} {et~al.}(1977){Weaver}, {McCray}, {Castor}, {Shapiro}, \&
  {Moore}}]{weaver_apj_218_1977}
{Weaver}, R., {McCray}, R., {Castor}, J., {Shapiro}, P., \& {Moore}, R. 1977,
  \apj, 218, 377

\bibitem[{{Weber} \& {Davis}(1967)}]{weber_apj_148_1967}
{Weber}, E.~J. \& {Davis}, Leverett, J. 1967, \apj, 148, 217

\bibitem[{{Whalen} {et~al.}(2008){Whalen}, {van Veelen}, {O'Shea}, \&
  {Norman}}]{whalen_apj_682_2008}
{Whalen}, D., {van Veelen}, B., {O'Shea}, B.~W., \& {Norman}, M.~L. 2008, \apj,
  682, 49

\bibitem[{{Wiersma} {et~al.}(2009){Wiersma}, {Schaye}, \&
  {Smith}}]{wiersma_mnras_393_2009}
{Wiersma}, R.~P.~C., {Schaye}, J., \& {Smith}, B.~D. 2009, \mnras, 393, 99

\bibitem[{{Wilkin}(1996)}]{wilkin_459_apj_1996}
{Wilkin}, F.~P. 1996, \apjl, 459, L31

\bibitem[{{Williams} {et~al.}(2013){Williams}, {Borkowski}, {Ghavamian},
  {Hewitt}, {Mao}, {Petre}, {Reynolds}, \& {Blondin}}]{williams_apj_770_2013}
{Williams}, B.~J., {Borkowski}, K.~J., {Ghavamian}, P., {et~al.} 2013, \apj,
  770, 129

\bibitem[{{Wolfire} {et~al.}(2003){Wolfire}, {McKee}, {Hollenbach}, \&
  {Tielens}}]{wolfire_apj_587_2003}
{Wolfire}, M.~G., {McKee}, C.~F., {Hollenbach}, D., \& {Tielens}, A.~G.~G.~M.
  2003, \apj, 587, 278

\bibitem[{{Woosley} {et~al.}(2002){Woosley}, {Heger}, \&
  {Weaver}}]{woosley_rvmp_74_2002}
{Woosley}, S.~E., {Heger}, A., \& {Weaver}, T.~A. 2002, Reviews of Modern
  Physics, 74, 1015

\bibitem[{{Woosley} \& {Weaver}(1986)}]{woosley_araa_24_1986}
{Woosley}, S.~E. \& {Weaver}, T.~A. 1986, \araa, 24, 205

\end{thebibliography}


\end{document}